\documentclass[acmsmall,screen]{acmart}
\usepackage{csquotes}
\usepackage{booktabs}
\usepackage{multirow}
\usepackage{tabularx}
\usepackage{adjustbox}
\usepackage{rotating}
\usepackage{enumitem}
\usepackage{nicematrix}
\usepackage{makecell}
\usepackage{color, colortbl}
\usepackage{tikz}
\usepackage{acronym}
\usepackage{placeins}
\newacro{TTIG}{Text-to-Image Generation}
\newacro{LLM}{Large Language Model}
\newacro{AI}{Artificial Intelligence}
\newacro{LTGM}{Large-scale Text-to-Image Generation Model}

\newcommand{\link}[1]{$\rightarrow$\textbf{#1}}
\newcommand{\rev}[1]{#1}

\AtBeginDocument{%
  \providecommand\BibTeX{{%
    \normalfont B\kern-0.5em{\scshape i\kern-0.25em b}\kern-0.8em\TeX}}}

\setcopyright{rightsretained}
\acmPrice{}
\acmDOI{10.1145/3611025}
\acmYear{2023}
\copyrightyear{2023}
\acmSubmissionID{play23main-p3812-p}
\acmJournal{PACMHCI}
\acmVolume{7}
\acmNumber{CHI PLAY}
\acmArticle{379}
\acmMonth{11}
\received{2023-02-21}
\received[accepted]{2023-07-07}

\begin{document}
\title[\enquote{An Adapt-or-Die Type of Situation}: Perception, Adoption, and Use of Text-to-Image-Generation AI (...)]{\enquote{An Adapt-or-Die Type of Situation}: Perception, Adoption, and Use of Text-to-Image-Generation AI by Game Industry Professionals}

\author{Veera Vimpari}
\email{veera.vimpari@aalto.fi}
\orcid{0000-0002-5591-6492}
\affiliation{%
  \institution{Aalto University}
  \department{Department of Computer Science}
  \city{Espoo}
  \country{Finland}
}

\author{Annakaisa Kultima}
\email{annakaisa.kultima@aalto.fi}
\orcid{0000-0001-5856-5643}
\affiliation{%
  \institution{Aalto University}
  \department{Department of Art \& Media}
  \city{Espoo}
  \country{Finland}
}

\author{Perttu Hämäläinen}
\email{perttu.hamalainen@aalto.fi}
\orcid{0000-0001-7764-3459}
\affiliation{%
  \institution{Aalto University}
  \department{Department of Computer Science}
  \city{Espoo}
  \country{Finland}
}

\author{Christian Guckelsberger}
\email{christian.guckelsberger@aalto.fi}
\orcid{0000-0003-1977-1887} 
\affiliation{
  \institution{Aalto University}
  \department{Department of Computer Science}
  \city{Espoo}
  \country{Finland}
}

\renewcommand{\shortauthors}{Veera Vimpari, Annakaisa Kultima, Perttu Hämäläinen, and Christian Guckelsberger}

\renewcommand{\authors}{Veera Vimpari, Annakaisa Kultima, Perttu Hämäläinen, and Christian Guckelsberger}

\begin{abstract}
Text-to-image generation (TTIG) models, a recent addition to creative AI, can generate images based on a text description. These models have begun to rival the work of professional creatives, and sparked discussions on the future of creative work, loss of jobs, and copyright issues, amongst other important implications. To support the sustainable adoption of TTIG, we must provide rich, reliable and transparent insights into how professionals perceive, adopt and use TTIG. Crucially though, the public debate is shallow, narrow and lacking transparency, while academic work has focused on studying the use of TTIG in a general artist population, but not on the perceptions and attitudes of professionals in a specific industry. In this paper, we contribute a qualitative, exploratory interview study on TTIG in the Finnish videogame industry. Through a Template Analysis on semi-structured interviews with 14 game professionals, we reveal 12 overarching themes, structured into 39 sub-themes on professionals' perception, adoption and use of TTIG in games industry practice. Experiencing (yet another) change of roles and creative processes, our participants' reflections can inform discussions within the industry, be used by policymakers to inform urgently needed legislation, and support researchers in games, HCI and AI to support the sustainable, professional use of TTIG, and foster games as cultural artefacts.
\end{abstract}



\maketitle

\section{Introduction}
\label{sec:introduction}

A fairly recent addition to \ac{AI}, \acf{TTIG} models such as DALL·E 2 \citep{openai_dalle2, ramesh2022hierarchical}, Midjourney \citep{midjourney}, and Stable Diffusion \citep{stable_diffusion, rombach2022high} enable the generation of images based on a text description, called the \emph{prompt}. Trained on huge datasets, these systems can produce high-quality images in a zero-shot fashion. \ac{TTIG} systems have begun rivalling the work of professional artists, and even won art prizes \citep{gault_2022}. They mark a major leap in the development of creative \ac{AI} more generally, not only characterised by unprecedented output quality, but also by an increase in creative imagination and autonomy \citep{jennings2010developing,colton2018issues}. Being arguably accessible and easy to use, these systems enable people to generate high-quality visuals without traditional artistic skills \cite{oppenlaender2022creativity}. Unsurprisingly, they have received much attention from users and the press. Already in Fall 2022, DALL·E 2 had 1.5 million active users \cite{openai_2022_1} and was used by many creative professionals across industries, from arts to architecture \citep{epstein2022happy}. Crucially though, \ac{TTIG} also provoked much controversy w.r.t. the future of creative work, loss of jobs, and copyright issues, amongst others. 

We hold that further advances on creative AI are not only inevitable, but welcome: we see a real opportunity for creative AI to be used in a sustainable fashion \citep{holzapfel2022environmental}, facilitating co-creation with professionals 
in a way that contributes to their artistic freedom and expression, self-realisation and well-being. However, to support this development, we must provide professionals, educators, policy makers and researchers with rich, reliable and transparent insights into how professionals perceive, adopt and use \ac{TTIG}. We are at a crossroads where this vital knowledge is largely missing \citep{moruzzi2020should}. As for AI at large \citep{floridi2021ethical}, we must study the good or harm that \ac{TTIG} is doing to its users.

While much has already been discussed by the press and on social media, the coverage is typically shallow, narrow, one-sided, and biased toward the often not clearly articulated agenda of the publishing body, thus offering little reliability to inform e.g.~policy development. On the academic side, only very few have studied the use of \ac{TTIG} systems by its users, while the majority of papers focus on technical contributions. 
Through an interview study of professional artists across many creative domains, \citet{ko2022large} have arguably presented the most impressive account of knowledge to date. Crucially though, their interviews and mostly deductive Thematic Analysis \citep{braun2006using} afford little exploration of unknown phenomena, but focus on the potential and weaknesses of current systems to support professionals in their work. In doing so, it betrays little information on peoples' attitudes and future projections toward these systems' impact on their job, practice and society more widely, i.e. precisely the controversial issues that are so important to address. This is complemented by Buraga's \citep{buraga2022emergence} netnography \citep{kozinets2014netnographic} study on how the Midjourney online community experience \ac{TTIG} in their creative process. Crucially though, the investigation rests on a mixed demographic of laypeople and professionals, within one community, and with respect to one system only. Based on this, it is unclear to which extent Buraga's findings apply to professional creatives specifically, and whether they generalise. \rev{Similarly, but concurrent with this study, \citet{inie2023} have investigating professionals attitudes through a qualitative, online survey study. Crucially though, their data collection and analysis is less exploratory, they consider creative \ac{AI} more generally beyond \ac{TTIG}, and their participants are not focused on a specific industry.} 
Given the high socioeconomic disruption potential of this technology, we argue that we must extend this work with studies of professionals in various, specific sub-demographics and in a more exploratory, fine-grained fashion to uncover and substantiate overlooked phenomena.

In this paper, we thus complement existing work through a qualitative, exploratory interview study of professional creatives' perception, adoption and use of \ac{TTIG} in one specific and localised demographic: the Finnish videogame industry. While professional creatives in games work mainly with digital tools, and are generally aware of and welcoming new technologies, their strong use of prototyping and ideation, as well as their product-oriented practice, also makes them arguably one of the most vulnerable groups w.r.t. \ac{TTIG}. \rev{This is critical, in that the Finnish game industry employs ca. 4100 people and, with a €3.2 billion turnover in 2021, is considered \enquote{a vital part of the national economy}~\citep{hiltunen_2022}.} Game industry specifically allows us to study professionals with varied backgrounds using \ac{TTIG} in a wide range of creative tasks, which are yet all geared toward one final product under commercial pressure. 
Given its pioneering role in the adoption of new technology, we believe that studying the perception, adoption and use of technology in game industry can provide important guidance for a sustainable development not only for other industries, but also individual artists. 
Next to our choice of demographic, we complement existing work through a more exploratory research paradigm, aiming to provide both deep and wide insights not only on professionals' present use, but also on their attitudes and future projections toward \ac{TTIG}. More specifically, we address the following three, overarching research questions:
\begin{enumerate}
    \item What are professionals' \textit{perceptions and attitudes} towards \ac{TTIG} systems and their future?
    \item How are \ac{TTIG} systems \textit{adopted and used} in the creative practice now?
    \item How will \ac{TTIG} systems change and be used in future creative practice?
\end{enumerate}
Through an in-depth, largely inductive, Template Analysis \citep{brooks2015utility}, we reveal 12 overarching themes in the interview data, structured into 39 sub-themes which provide substantially richer and more detailed insights compared to related work, but also complement it. We pair a detailed description of all themes with a summary discussion w.r.t. our research questions, and thus contribute the most comprehensive qualitative user study on \ac{TTIG} to date. In contrast to related work \citep{ko2022large}, our interviews were conducted in English, and our comprehensive report with many quotes (Supplementary) can thus communicate rich impressions of our findings to a wider audience. While all our participants are working in game industry, we crucially understand our target demographic -- professional creatives -- not primarily as employees but as people who work creatively at a high standard, typically informed by a formal education in a creative subject. 
As our study will show, such professional use is however not bound to the workplace. 
Thus, while not the main focus, our study also highlights interactions in the use of \ac{TTIG} between these contexts. Finally, our study contributes urgently needed insights on professionals' first-hand experiences when interacting with creative AI more generally.






\section{Background: Text-to-Image Generation}
\label{sec:background}

\begin{figure}[t!]
    \centering
    \includegraphics[width=\textwidth]{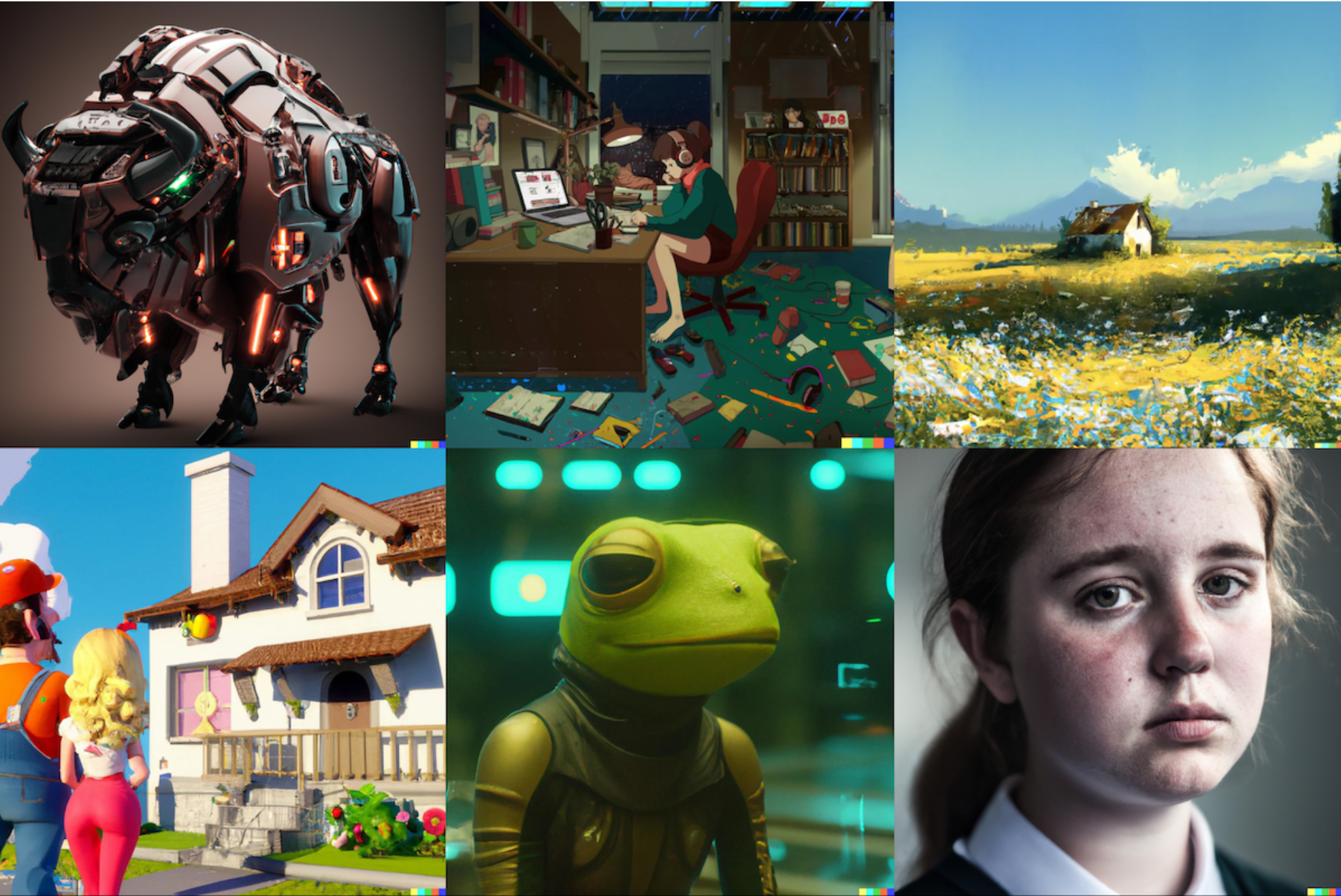}
    \caption{Example images generated with OpenAI's DALL·E 2, demonstrating a range of styles and subjects relevant to videogame production. The images were taken from the r/dalle2 subreddit, links to original posts corresponding to images from the top-left to bottom-right: \href{http://bit.ly/3Un8e5k}{1}, \href{http://bit.ly/3TpWq0V}{2}, \href{http://bit.ly/3TkxyHH}{3}, \href{http://bit.ly/3Uot1W3}{4}, \href{http://bit.ly/3Umn9wJ}{5}, \href{http://bit.ly/3hvXVxi}{6}. The same examples were used in our study invitation (Sec.~\ref{sec:study_recruitment}).}
    \label{fig:ttig_example_images}
\end{figure}

\acf{TTIG} systems, also referred to as \acp{LTGM}, allow for the generation of images based on a text description, called the \emph{prompt}. \rev{We provide several examples of such images in Fig.~\ref{fig:ttig_example_images}. The caption moreover comprises links to the prompts used to generate them and their authors. For instance, the image in the top-left corner was produced with the prompt:
\begin{quote}
    a cybertronic bison, leds, high detail, sharp, studio, digital art
\end{quote}}
\rev{While our example is very compact, prompts can become long and complex when aiming to generate a specific subject or style.}
\ac{TTIG} systems represent a recent addition to \ac{AI}, supported by the introduction of \acp{LLM} \citep{brown2020language}. The latter are capable of generating highly realistic text responses to a range of specialised tasks such as question answering or classification, based on little (few-shot learning) or no prior examples (zero-shot learning). In tandem with the development of \acp{LLM}, \citet{mansimov2015generating} demonstrated that generative image models conditioned on image captions can generate images from natural language input. Combined with LLMs, \ac{TTIG} models are taught to associate text descriptions with visual attributes, and to synthesise images corresponding to natural language inputs.

In January 2021, OpenAI released the first version of DALL·E \citep{openai_dalle, ramesh2021zero}. The system combines a Transformer \citep{vaswani2017attention} architecture with Contrastive Language-Image Pre-training \citep[CLIP,][]{ramesh2022hierarchical}, a separate model which has been trained on large amounts of image and image caption data to learn the relationships between image features and their description. 
\ac{TTIG} systems became a mainstream phenomenon in Spring 2022 when DALL·E's successor DALL·E 2 \citep{openai_dalle2, ramesh2022hierarchical} was released. Replacing the Transformer with a Diffusion \citep{ho2020denoising} architecture, amongst other changes, enabled the generation of more realistic and accurate images. Shortly after, other state-of-the-art systems such as Midjourney \citep{midjourney}, and the open-source model Stable Diffusion \citep{stable_diffusion, rombach2022high} were released, fuelling the popularity of \ac{TTIG}. \rev{We detail the version history of the most popular \ac{TTIG} systems up to the period of conducting our interviews in Tbl.~\ref{tbl:ttig_versions}.}
\begin{figure}[b!]
    \centering
    \includegraphics[width=\linewidth]{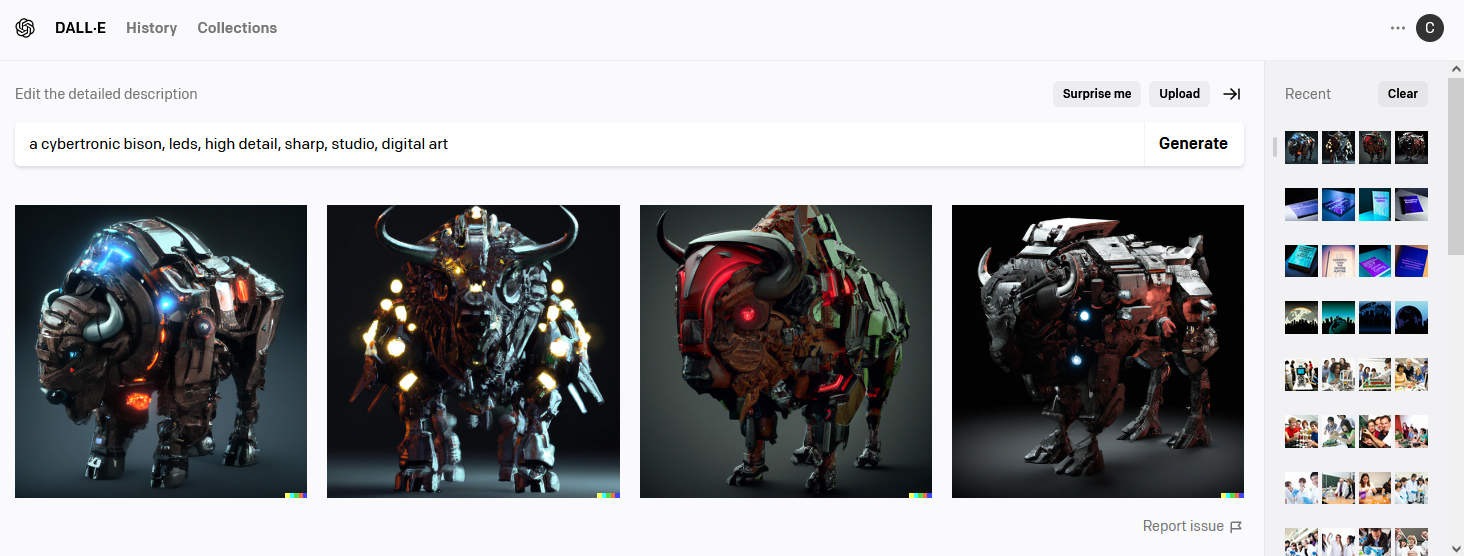}
    \caption{\rev{The DALL·E 2 web interface (June 2023) with four variations of the top-left image in Fig.~\ref{fig:ttig_example_images}, generated with the same prompt. The original image from our example is not present, highlighting the stochastic nature of the generation process.}}
    \label{fig:dalle_2_interface}
\end{figure}%
Most of these systems can be accessed without a separate installation, e.g. DALL·E 2 \citep{openai_dalle2, ramesh2022hierarchical} and Midjourney \citep{midjourney} online and through Discord, respectively (Fig.~\ref{fig:dalle_2_interface}). They operate on a freemium subscription model, meaning that the user can generate a number of images for free, but must pay for additional images or features. Other systems such as Stable Diffusion \citep{stable_diffusion} are only available online through a paid subscription, but can be installed locally free of charge through an installation process which various initiatives have successively made more approachable by non experts.
\begin{table}[b!]
    \centering
    \caption{\rev{Version history of the three \ac{TTIG} systems that were most popular with our participants, up to and including the interview period (November-December 2022) and only including official releases.}}
    \begin{tabularx}{\textwidth}{Xl}
        \toprule
        System & Version\\
        \midrule
        DALL·E & V1 (01/21), V2 with wait list (07/22), V2 open (09/22)\\
        Stable Diffusion & V1 (08/22), V2 (11/22), V2.1 (12/22)\\
        Midjourney & V1 (02/22), V2 (04/22), V3 (07/22), V4 (09/22)\\
        \bottomrule 
    \end{tabularx}
    \label{tbl:ttig_versions}
\end{table}
Most \ac{TTIG} systems are arguably accessible and easy to use, enabling the user to generate high-quality visual content without traditional artistic skills \cite{oppenlaender2022creativity}. Unsurprisingly, they have received a lot of attention from both, users and the press. Already in September 2022, DALL·E 2 had 1.5 million active users and is used by many creative professionals across industries, from arts to architecture \cite{openai_2022_1}. Crucially though, \ac{TTIG} systems also have brought up a lot of controversy \citep{holzapfel2022environmental}. We shed more light into the debate through the following survey of related work.


\section{Related Work}
\label{sec:related_work}

\ac{TTIG} systems have provoked widespread attention on different internet forums and social media platforms, and they caught the interest of popular news outlets. At the same time, researchers have responded much more slowly, likely due to the effort required for deeper study, and the time spent in peer-reviewed publication cycles. We consequently complement a short review of academic related work (Sec.~\ref{sec:related_academic}) with a longer survey of non-academic publications (Sec.~\ref{sec:related_non_academic}), yielding a priori codes and themes for our analysis. We relate our study findings back to related work in Sec.~\ref{sec:related_work_comparison}.


\subsection{Academic Related Work}
\label{sec:related_academic}

At the time of submitting this paper, only two academic studies have investigated the use of \ac{TTIG} systems through a user study specifically, and only one study specifically on a population of professional creatives. 

Interested in how visual artists adopt \ac{TTIG} systems to support their creative practice, \citet{ko2022large} conducted semi-structured interviews with 28 visual artists working in 35 visual art domains, most aged between 20 and 30. The artists were familiarised with one specific system (DALL·E), and subjected to a semi-structured interview in their native Korean tongue, and grounded in the three topics of user, task, and role. They applied Thematic Analysis to analyse this data, primarily deductively based on their topics, and complemented with inductive analysis, albeit within the theoretical frame of their research questions. They structure their findings based on the potential benefits \ac{TTIG} systems offer to visual artists, and their current limitations. More specifically, they find that \ac{TTIG} systems can benefit artists (1) as image reference search tools; (2) enable visual communication; (3) rectify people's biased creation (i.e. inspire out-of-the-box thinking); (4) bring fast prototyping to novice artists; 5) and lend justification to \enquote{a New Era of AI Art}. Moreover, they find that \ac{TTIG} systems at present limit artists in that (1) they can only generate predictable images; (2) do not support personalisation; (3) restrain creativity through the prompting mechanism; and (4) are inefficient and become a burden. They conclude that visual artists found it hard to actually incorporate \ac{TTIG} into their creative works in its current form, and put forward four design guidelines to improve \ac{TTIG} user interfaces and experience. We complement this study by focusing on creatives in the Finnish game industry as one specific professional demographic. We include more experienced professionals, also in lead roles, and highlight differences in perception between those who direct, and those who approach creative work first hand. Moreover, we go beyond use and also consider people's attitudes and future projections toward these systems, including ethical issues and societal impact. By analysing the data in a mostly inductive fashion, we support the discovery of previously unknown phenomena. Finally, our interviews were conducted in English, and we complement our summary in Sec.~\ref{sec:results} with an extensive and accessible report of our findings illustrated with the original quotes (Supplementary).

\citet{buraga2022emergence} investigated how the Midjourney online community experience the AI system in their creative process, and which topics the members discussed most. To this end, they employed Netnography \cite{kozinets2014netnographic}, a qualitative research method which adopts ethnography to the context of modern media platforms, to analyse a total of 300 social media posts from Twitter and Facebook. 
Buraga finds that the Midjourney users perceive the AI as a collaborative tool which supports and augments their creativity, or even becomes an essential part of their creative process. The community members understand their role to transform into that of a commissioner, but also note that creating art with \ac{TTIG} systems is still hard work. Some community members considered AI-created art to be a new form of art, and some were critical of AI art generators, due to fear of replacing artists' jobs. In also embracing attitudes toward \ac{AI} use, \citet{buraga2022emergence} can be seen to complement \citet{ko2022large}'s work. However, it is not clear to which extent their findings generalise, given the focus on a single community and \ac{TTIG} system. In addition, the Midjourney community consists of both professionals and laypeople, and we can thus not delimit which insights pertain to professional creatives specifically. In contrast, our study focuses on professional creatives in a single industry, using any form of \ac{TTIG} system. Moreover, our participants likely developed their views and practices more independently from a single community. 

\rev{\citet{inie2023} have published a related, work-in-progress study after the original submission of this manuscript. Through a qualitative, online survey with six open-ended questions, they asked 23 participants from six countries about their definition of AI, beliefs on whether AI can be creative, worries and excitement about AI contributing to creative work, and the expected role of creative AI in the near and far future. Through a mostly deductive Thematic Analysis, they identify four themes closely aligned with their research questions, and structured into nine sub-themes. The participants' concerns are reflected in three sub-themes: (1) worse quality output (and risk of job loss); (2) weakening the creative process; and (3) copyright issues. This is complemented by three sub-themes capturing reasons for not worrying about creative AI: (1) AI cannot produce output without human input; (2) AI output is not convincing; (3) my work/creative process is too complex for AI to imitate. A third theme captured participants' excitement about creative AI, structured into the sub-themes (1) AI can raise productivity for the individual or for larger processes; (2) AI can offer inspiration; and (3) AI can lead to higher quality output. They discuss how their findings could empower professional creatives by informing the design of participatory AI \citep{Abeba2022}. We applaud this study for its focus on people's feelings and attitudes. Moreover, by proposing a specific three-component definition of creativity, the authors gain rich insights into the attributes people use when (not) attesting creativity to AI. Crucially though, the findings are less specific than the ones presented here, in that the study did not explicitly ask for thoughts on \ac{TTIG} systems, but (creative) AI more generally. In not addressing adoption and use, the study is also comparatively less holistic. Moreover, the authors invited self-defining creative professionals from different industries who deem creativity to \enquote{play a significant role in their work}; consequently, their participants comprise various professions such as software engineer, librarian and researchers, which we excluded with our definition (Sec.~\ref{sec:introduction}). The authors admit that, coming from different professions, the participants' \enquote{everyday work lives may (...) not necessarily be impacted in the same ways by generative AI}. Consequently, they call for future work to \enquote{categorize different creative industries and identify which and how specific work tasks within these industries may be impacted by generative AI}. Here, we contribute to this agenda by investigating the perception, adoption and use of \ac{TTIG} in the game industry.}


\subsection{Non-Academic Related Work}
\label{sec:related_non_academic}



We complement these sparse academic insights\footnote{The study by \citet{ko2022large}, arguably a rich and the most relevant piece of related work, had only been released after the time of conducting this survey. Moreover, the study by \citet{inie2023} has been published after submitting the first version of this manuscript.} 
with a non-exhaustive survey of %
non-academic articles which voice the views of professional creatives on \ac{TTIG} systems. \rev{Based on this survey, we identify a priori codes to connect our exploratory study (Sec.~\ref{sec:study}) with the existing debate}. The survey was conducted between August and October 2022.

To support the accuracy and trustworthiness of our findings, we focused on news articles and artist interviews by professional reporters on reputable news sites and in digital publications. We did not include, for example, social media posts, discussion forums, private blogs, and comments from independent video creators. More specifically, we used Google search with queries combining \enquote{text-to-image generation} with system names (\enquote{DALL·E}, \enquote{DALL·E 2}, \enquote{Midjourney}, \enquote{Stable Diffusion}) and the terms \enquote{artist}, \enquote{professional creative}, \enquote{react}, \enquote{interview}, and \enquote{collaborate}. We screened all publications within the first 100 hits per query. We only included original articles to avoid duplicates and repetition. Moreover, we excluded articles which did not directly present the voice of a professional creative, e.g.~instances of industry announcements. This filtering processes resulted in a final selection of 22 articles. The articles were analysed by authors 1 and 4 through inductive, non-hierarchical coding \citep{crabtree1992template} but a theoretical focus on our research questions (Sec.~\ref{sec:introduction}). The coding fragment was the entire article, and we assigned multiple codes to each article.

Overall, we found 11 topics, and their assignment to each code is summarised in Tbl.~\ref{tbl:related_non_academic}. Next, we briefly summarise each of these topics, supported by quotes. We ordered them not according to prevalence, but to support the resolution of dependencies between topics.

\paragraph{Creative process:} Many sources focus on the nature of the creative process when \ac{TTIG} systems are incorporated. Most frequently, it is described as iterative, i.e.~the text prompt is tweaked and fine-tuned continuously to achieve the desired results \citep{harwell_2022, hertzmann_2022, openai_2022, rajnerowicz_2022, droitcour_2022}. The process also often includes post-editing with software like Adobe Photoshop, as the systems still lack, e.g., in generating certain compositions, or human hands and faces \citep{harwell_2022, gopani_2022, gault_2022}. Some articles emphasise the exploration and combination of ideas, playing with the system, and pushing its boundaries \citep{vox.com_2022, furman_2022}.

\paragraph{Roles in the creative process:} Related to the creative process more generally, many articles discuss the roles which the human creative and the AI system assume. Some articles only question who is the artist when \ac{TTIG} systems become involved \cite{vox.com_2022, furman_2022}. Other articles assign the human artist the role of curating the system's outputs \cite{furman_2022, ibrahim_2022, harwell_2022}. Related, some stress that, when using \ac{TTIG} systems, artists take more responsibility for the ideas and concepts they create, thus assuming the role of an art director \cite{ibrahim_2022, goldman_2022, vox.com_2022}. Nevertheless, according to many sources, the role of the person's artistic background is still essential in initiating and guiding the creative process. The systems are conceived as tools or materials \cite{openai_2022, karmaker_2022}, tireless team members \cite{ibrahim_2022} or assistants \cite{knight_2022} that feed the artist endless ideas to work with:
\begin{quote}
    (...) I also like having an inert member of \enquote{the team} who just spits shit out.\\-- Richard Turley, quoted by Alif Ibrahim, It's Nice That \cite{ibrahim_2022}
\end{quote}

\paragraph{Using output as a source of inspiration:} Many articles emphasise the exploratory use of \ac{TTIG} systems, yielding surprising outputs which can spark novel ideas and boost the artist's creativity \cite{hertzmann_2022, knight_2022, diaz_2022, rajnerowicz_2022, openai_2022}:
\begin{quote}
    We didn’t know what an osteosarcoma villain would look like so we turned to DALL·E as our creative outlet. DALL·E gave us a huge amount of inspiration.\\-- James and Kristin Orrigo, quoted by OpenAI \cite{openai_2022}
\end{quote}
Another source suggests that such engagement with the systems can enable artists in exploring their own identity \cite{lawson-tancred_2022}:
\begin{quote}
    The conjuring of thoughts into visuals enables the exploration of creativity. It can be a powerful tool for new artists to figure out that they are artists.\\-- Apolinário Passos, quoted by Jo Lawson-Tancred, Artnet \cite{lawson-tancred_2022}
\end{quote}
In some articles, the use of \ac{TTIG} is compared to online image search for reference pictures and inspiration, a common practice in creative work \cite{furman_2022, ibrahim_2022, goldman_2022}.

\paragraph{Prototyping/Conceptualising ideas} Many articles highlight that \ac{TTIG} systems can be helpful in conceptualising ideas \cite{rajnerowicz_2022, goldman_2022}. Artists can, for example, generate multiple versions of an idea to quickly communicate to a client which version of a concept to pursue \cite{goldman_2022}. Also, the systems allow those without artistic skills to bring their concepts to life \cite{openai_2022}.
\begin{quote}
    For enterprise clients, this technology can provide a vehicle to get from idea to concept and then help to refine the concept much faster.\\-- Andy Martinus, quoted by Sharon Goldman, Venture Beat \cite{goldman_2022}
\end{quote}

\paragraph{Increased efficiency} DALL·E 2 and other \ac{TTIG} systems been argued to perform tasks that used to take days now within seconds. Some sources arge that automating tasks in the design process can free time for artists to be spent on more gratifying tasks such as ideation, creative decision-making, and exploring new techniques \cite{ibrahim_2022, karmaker_2022, diaz_2022, flowers_2022}:
\begin{quote}
Automation has always been interesting to me simply because I enjoy being creative, but I don’t like to execute it. I like to get things done as fast as possible.\\-- Daniel Wenzel quoted by Alif Ibrahim, It's Nice That \cite{ibrahim_2022}
\end{quote}
Others have highlighted potential money savings in production due to such automation \cite{goldman_2022, rajnerowicz_2022}.

\newcommand{\tikzcircle}[2][black, fill=black]{\tikz[baseline=-0.5ex]\draw[#1,radius=#2] (0,0) circle ;}%
\newcommand{\tikzcircleempty}[2][black, fill=none]{\tikz[baseline=-0.5ex]\draw[#1,radius=#2] (0,0) circle ;}%
\definecolor{Gray}{gray}{0.9}
\begin{table}[t!]
\small
\centering
\caption{The 22 non-academic articles identified through our search, distinguished by their coding (black circle) on 11 topics on professional creatives' perception, adoption and use of \ac{TTIG} systems, identified in the review process.}
\begin{adjustbox}{max width=\textwidth}
\begin{tabular}{l c c c c c c c c c c c c c}
\textbf{Title (abbreviated)} & \begin{turn}{90} Creative Process\end{turn} & \begin{turn}{90} Roles in the creative process\end{turn} & \begin{turn} {90} Prototyping/conceptualising ideas \end{turn} & \begin{turn}{90} Using output as a source of inspiration\end{turn} & \begin{turn}{90} Increased efficiency \end{turn} & \begin{turn}{90} Artists' jobs \end{turn} &\begin{turn}{90} Democratisation of creative work \end{turn} &\begin{turn}{90} Devaluation of creative work \end{turn} & \begin{turn}{90} Bias and harmful content \end{turn} & \begin{turn}{90} Copyright issues \end{turn} & \begin{turn}{90} Big tech critique \end{turn} \\
\hline

He used AI to win a fine-arts competition... \cite{harwell_2022} & \tikzcircle{2pt} & 	\tikzcircle{2pt} & 	 $\tikzcircleempty{2pt}$ &	 $\tikzcircleempty{2pt}$ &	 $\tikzcircleempty{2pt}$ &	 $\tikzcircleempty{2pt}$ &	 $\tikzcircleempty{2pt}$ &	 $\tikzcircleempty{2pt}$ &	 $\tikzcircleempty{2pt}$ &	\tikzcircle{2pt} & 	 $\tikzcircleempty{2pt}$ \\

\rowcolor{Gray}
When AI Makes Art, Humans Supply the Creative Spark \cite{knight_2022} & \tikzcircle{2pt} & 	 $\tikzcircleempty{2pt}$ &	 $\tikzcircleempty{2pt}$ &	 $\tikzcircleempty{2pt}$ &	\tikzcircle{2pt} & 	\tikzcircle{2pt} & 	\tikzcircle{2pt} & 	 $\tikzcircleempty{2pt}$ &	\tikzcircle{2pt} & 	\tikzcircle{2pt} & 	 $\tikzcircleempty{2pt}$ \\

Give this AI a few words of description... \cite{hertzmann_2022}  & \tikzcircle{2pt} & 	 $\tikzcircleempty{2pt}$ &	 $\tikzcircleempty{2pt}$ &	\tikzcircle{2pt} & 	 $\tikzcircleempty{2pt}$ &	 $\tikzcircleempty{2pt}$ &	 $\tikzcircleempty{2pt}$ &	 $\tikzcircleempty{2pt}$ &	\tikzcircle{2pt} & 	 $\tikzcircleempty{2pt}$ &	 $\tikzcircleempty{2pt}$ \\

\rowcolor{Gray}
‘A.I. Is Being Trained on the Collective Works of Humanity’ \cite{lawson-tancred_2022} &  $\tikzcircleempty{2pt}$ &	 $\tikzcircleempty{2pt}$ &	 $\tikzcircleempty{2pt}$ &	\tikzcircle{2pt} & 	 $\tikzcircleempty{2pt}$ &	 $\tikzcircleempty{2pt}$ &	\tikzcircle{2pt} & 	 $\tikzcircleempty{2pt}$ &	 $\tikzcircleempty{2pt}$ &	\tikzcircle{2pt} & 	 $\tikzcircleempty{2pt}$ \\

The Future of AI-Generated Art Is Here... \cite{flowers_2022} & \tikzcircle{2pt} & 	 $\tikzcircleempty{2pt}$ &	 $\tikzcircleempty{2pt}$ &	 $\tikzcircleempty{2pt}$ &	\tikzcircle{2pt} & 	\tikzcircle{2pt} & 	 $\tikzcircleempty{2pt}$ &	 $\tikzcircleempty{2pt}$ &	 $\tikzcircleempty{2pt}$ &	 $\tikzcircleempty{2pt}$ &	 $\tikzcircleempty{2pt}$ \\

\rowcolor{Gray}
Bonus video: What AI art means for human artists \cite{vox.com_2022} & \tikzcircle{2pt} & 	\tikzcircle{2pt} & 	 $\tikzcircleempty{2pt}$ &	\tikzcircle{2pt} & 	 $\tikzcircleempty{2pt}$ &	\tikzcircle{2pt} & 	 $\tikzcircleempty{2pt}$ &	\tikzcircle{2pt} & 	 $\tikzcircleempty{2pt}$ &	\tikzcircle{2pt} & 	 $\tikzcircleempty{2pt}$ \\

From ‘Barbies scissoring’ to ‘contorted emotion’... \cite{furman_2022} & \tikzcircle{2pt} & 	\tikzcircle{2pt} & 	 $\tikzcircleempty{2pt}$ &	\tikzcircle{2pt} & 	 $\tikzcircleempty{2pt}$ &	 $\tikzcircleempty{2pt}$ &	 $\tikzcircleempty{2pt}$ &	 $\tikzcircleempty{2pt}$ &	\tikzcircle{2pt} & 	 $\tikzcircleempty{2pt}$ &	 $\tikzcircleempty{2pt}$ \\

\rowcolor{Gray}
DALL·E 2: Extending Creativity \cite{openai_2022} & \tikzcircle{2pt} & 	\tikzcircle{2pt} & 	\tikzcircle{2pt} & 	 $\tikzcircleempty{2pt}$ &	 $\tikzcircleempty{2pt}$ &	 $\tikzcircleempty{2pt}$ &	 $\tikzcircleempty{2pt}$ &	 $\tikzcircleempty{2pt}$ &	 $\tikzcircleempty{2pt}$ &	 $\tikzcircleempty{2pt}$ &	 $\tikzcircleempty{2pt}$ \\

Algorithms Can Now Mimic Any Artist. Some Artists Hate It \cite{knight_2022_2} &  $\tikzcircleempty{2pt}$ &	 $\tikzcircleempty{2pt}$ &	 $\tikzcircleempty{2pt}$ &	 $\tikzcircleempty{2pt}$ &	 $\tikzcircleempty{2pt}$ &	 $\tikzcircleempty{2pt}$ &	 $\tikzcircleempty{2pt}$ &	 $\tikzcircleempty{2pt}$ &	\tikzcircle{2pt} & 	\tikzcircle{2pt} & 	\tikzcircle{2pt}  \\

\rowcolor{Gray}
Machine Learning: Are designers even needed anymore? \cite{ibrahim_2022} & \tikzcircle{2pt} & 	\tikzcircle{2pt} & 	 $\tikzcircleempty{2pt}$ &	\tikzcircle{2pt} & 	\tikzcircle{2pt} & 	\tikzcircle{2pt} & 	 $\tikzcircleempty{2pt}$ &	 $\tikzcircleempty{2pt}$ &	 $\tikzcircleempty{2pt}$ &	\tikzcircle{2pt} & 	 $\tikzcircleempty{2pt}$ \\

An AI-Generated Artwork Won First Place... \cite{gault_2022} & \tikzcircle{2pt} & 	 $\tikzcircleempty{2pt}$ &	 $\tikzcircleempty{2pt}$ &	 $\tikzcircleempty{2pt}$ &	 $\tikzcircleempty{2pt}$ &	\tikzcircle{2pt} & 	 $\tikzcircleempty{2pt}$ &	 $\tikzcircleempty{2pt}$ &	 $\tikzcircleempty{2pt}$ &	 $\tikzcircleempty{2pt}$ &	\tikzcircle{2pt} \\

\rowcolor{Gray}
What do the artists think about AI image generators? \cite{karmaker_2022} & \tikzcircle{2pt} & 	\tikzcircle{2pt} & 	 $\tikzcircleempty{2pt}$ &	 $\tikzcircleempty{2pt}$ &	\tikzcircle{2pt} & 	\tikzcircle{2pt} & 	 $\tikzcircleempty{2pt}$ &	 $\tikzcircleempty{2pt}$ &	 $\tikzcircleempty{2pt}$ &	 $\tikzcircleempty{2pt}$ &	 $\tikzcircleempty{2pt}$ \\

The trouble with DALL·E \cite{droitcour_2022} & \tikzcircle{2pt} & 	 $\tikzcircleempty{2pt}$ &	 $\tikzcircleempty{2pt}$ &	 $\tikzcircleempty{2pt}$ &	 $\tikzcircleempty{2pt}$ &	 $\tikzcircleempty{2pt}$ &	 $\tikzcircleempty{2pt}$ &	 $\tikzcircleempty{2pt}$ &	 $\tikzcircleempty{2pt}$ &	\tikzcircle{2pt} & 	\tikzcircle{2pt} \\



\rowcolor{Gray}
This 30-second fashion show demonstrates... \cite{diaz_2022} &  $\tikzcircleempty{2pt}$ &	 $\tikzcircleempty{2pt}$ &	 $\tikzcircleempty{2pt}$ &	 $\tikzcircleempty{2pt}$ &	\tikzcircle{2pt} & 	\tikzcircle{2pt} & 	\tikzcircle{2pt} & 	 $\tikzcircleempty{2pt}$ &	 $\tikzcircleempty{2pt}$ &	 $\tikzcircleempty{2pt}$ &	 $\tikzcircleempty{2pt}$ \\

OpenAI’s new image generator sparks both excitement and fear \cite{macaulay_2022} &  $\tikzcircleempty{2pt}$ &	 $\tikzcircleempty{2pt}$ &	 $\tikzcircleempty{2pt}$ &	\tikzcircle{2pt} & 	 $\tikzcircleempty{2pt}$ &	\tikzcircle{2pt} & 	 $\tikzcircleempty{2pt}$ &	 $\tikzcircleempty{2pt}$ &	\tikzcircle{2pt} & 	 $\tikzcircleempty{2pt}$ &	 $\tikzcircleempty{2pt}$ \\

\rowcolor{Gray}
Will OpenAI's DALL·E kill creative careers? \cite{goldman_2022} &  $\tikzcircleempty{2pt}$ &	\tikzcircle{2pt} & 	\tikzcircle{2pt} & 	 $\tikzcircleempty{2pt}$ &	\tikzcircle{2pt} & 	\tikzcircle{2pt} & 	 $\tikzcircleempty{2pt}$ &	\tikzcircle{2pt} & 	 $\tikzcircleempty{2pt}$ &	\tikzcircle{2pt} & 	 $\tikzcircleempty{2pt}$ \\

Does DALL·E pose a threat to designer jobs? \cite{gopani_2022} & \tikzcircle{2pt} & 	 $\tikzcircleempty{2pt}$ &	 $\tikzcircleempty{2pt}$ &	 $\tikzcircleempty{2pt}$ &	 $\tikzcircleempty{2pt}$ &	\tikzcircle{2pt} & 	\tikzcircle{2pt} & 	\tikzcircle{2pt} & 	 $\tikzcircleempty{2pt}$ &	 $\tikzcircleempty{2pt}$ &	 $\tikzcircleempty{2pt}$ \\

\rowcolor{Gray}
Will AI \ac{TTIG} systems Turn Us All Into Artists? \cite{rajnerowicz_2022} & \tikzcircle{2pt} & 	 $\tikzcircleempty{2pt}$ &	\tikzcircle{2pt} & 	\tikzcircle{2pt} & 	\tikzcircle{2pt} & 	 $\tikzcircleempty{2pt}$ &	 $\tikzcircleempty{2pt}$ &	 $\tikzcircleempty{2pt}$ &	\tikzcircle{2pt} & 	\tikzcircle{2pt} & 	 $\tikzcircleempty{2pt}$ \\

Plagiarism by Machine \cite{savov_2022} &   $\tikzcircleempty{2pt}$ &	 $\tikzcircleempty{2pt}$ &	 $\tikzcircleempty{2pt}$ &	 $\tikzcircleempty{2pt}$ &	 $\tikzcircleempty{2pt}$ &	\tikzcircle{2pt} & 	 $\tikzcircleempty{2pt}$ &	 $\tikzcircleempty{2pt}$ &	 $\tikzcircleempty{2pt}$ &	\tikzcircle{2pt} & 	\tikzcircle{2pt} \\


\rowcolor{Gray}
This artist is dominating AI-generated art... \cite{heikkilä_2022} &  $\tikzcircleempty{2pt}$ &	 $\tikzcircleempty{2pt}$ &	 $\tikzcircleempty{2pt}$ &	 $\tikzcircleempty{2pt}$ &	 $\tikzcircleempty{2pt}$ &	\tikzcircle{2pt} & 	 $\tikzcircleempty{2pt}$ &	 $\tikzcircleempty{2pt}$ &	 $\tikzcircleempty{2pt}$ &	\tikzcircle{2pt} & 	 $\tikzcircleempty{2pt}$ \\
\end{tabular}
\end{adjustbox}
\label{tbl:related_non_academic}
\end{table}

\paragraph{Impact on creatives’ jobs} As automation increases efficiency in creative practice, it is natural to ask whether and how this will impact creative jobs. This question indeed has been raised by the majority of our sources, with some provocatively asking whether creative people will be needed anymore at all, or be replaced by AI \cite{ibrahim_2022, gopani_2022}. The reviewed sources particularly name artists \cite{gault_2022, knight_2022_2}, illustrators \cite{heikkilä_2022, vox.com_2022}, junior designers \cite{ibrahim_2022}, commercial designers \cite{macaulay_2022}, and photographers\cite{savov_2022} as those under threat. Also, the stock-photo industry is predicted to lose in relevance, as \ac{TTIG} systems could be used to create custom stock imagery \cite{vox.com_2022, gopani_2022}. 
This pessimistic view is opposed by more optimistic articles, suggesting that \ac{TTIG} woud not eliminate jobs, bur create new ones \cite{goldman_2022}. They typically remind of the nature of technological progress and historical instances in which technology has changed creative professions, created new ones, and eliminated some \cite{macaulay_2022, goldman_2022}. Most prominently, many compare the development of \ac{TTIG} to that of photography, first evoking fear and dismissal, but ultimately developing into its own art form and demand of new expertise \cite{vox.com_2022, harwell_2022, knight_2022}. 
One of the most evident new roles that \ac{TTIG} has created is the prompt engineer \cite{lawson-tancred_2022}, prompt designer \cite{flowers_2022}, or prompt crafter \cite{vox.com_2022} who focuses on tuning the text prompt for \ac{TTIG} to optimise a desired output. Even though technically, most people can write such prompts, some highlight that deeper knowledge of arts and design is required, e.g. to include fitting references to historical styles. Optimising an image to a high standard is argued to take time and many iterations, which the majority of users are likely not willing to devote \cite{lawson-tancred_2022, vox.com_2022} As a consequence, prompts have become a currency, and many artists refuse to share their prompts to keep the value to themselves \cite{harwell_2022}. With creative careers predicted to be in flux, some sources highlight the need for artists to acquire the skills for using \ac{TTIG} to keep up with the change \cite{vox.com_2022, diaz_2022, rajnerowicz_2022}. Some suggest that the value of ideas and creativity will increase, implying that particularly artists who are skilled in their craft but have limited creativity, are at risk of losing their jobs \cite{diaz_2022}:
\begin{quote}
    I can see how truly creative people will be safe and empowered by these new tools rather than threatened by them. I can imagine people who are great at using After Effects or Photoshop, but have limited creativity, losing jobs, the same way that many other jobs were lost to technology that empowered others to do amazing things. -- Paul Trillo, quoted by Jesus Diaz, Fast Company \cite{diaz_2022}
\end{quote}

\paragraph{Democratisation of creative work} Many argue that \ac{TTIG} makes digital art creation more accessible, and see a positive impact in anyone being able to express their ideas visually \cite{diaz_2022}:
\begin{quote}
    In one sense, AI democratizes image-making so that people who are more verbal can express themselves visually. -- Paul Trillo, quoted by Jesus Diaz, Fast Company \cite{diaz_2022}
\end{quote}
This line of thought also has implications in a professional context: several sources argue that, when art creation is no longer available only to the ones with the required skills and resources, teams with smaller budgets can introduce more visual content in their production \cite{lawson-tancred_2022, knight_2022, goldman_2022, diaz_2022}.

\paragraph{Devaluation of creative work} As a consequence of \ac{TTIG}'s accessibility, some sources predict that the digital art space can become even more saturated \cite{knight_2022}. Related, others argue that the democratisation of image production will decrease the value of creative work 
\cite{gopani_2022, vox.com_2022}. Furthermore, the ability to generate high-quality visuals at a low cost could reduce the demand for truly outstanding art, which in turn could lead to increased competition among the most skilled professional creatives, pushing the price of creative labour even lower \cite{goldman_2022}. However, some also project that, as the value of artistic work goes down, good ideas will be valued more \cite{vox.com_2022}.

\paragraph{Copyright issues} One of the most discussed topics concerns copyright. The controversy is twofold: Firstly, do artists have a say in, and get compensated for the use of their work as \ac{TTIG} training data? Many resent that their artwork has been used in training \ac{TTIG} systems without their consent, especially because adding their name to the text prompt then allows anyone to create images in their style without granting them any credit or compensation \cite{heikkilä_2022, savov_2022, knight_2022_2}:
\begin{quote}
    My objection is that they make plagiarism trivial and easy: you just feed them my photos, combine them with the vast trove of other people’s images you scrape off the web, and they’ll spit out any scene you like in my particular style. And now that they’re going commercial — DALL·E’s price is 15 for 115 text-based prompts — they’ll completely hollow out the already emaciated market for professional photography, editing, and imagery. -- Vlad Savov, Bloomberg \cite{savov_2022}
\end{quote}
\begin{quote}
     AI is the latest and most vicious of these technologies. It basically takes lifetimes of work by artists, without consent, and uses that data as the core ingredient in a new type of pastry that it can sell at a profit with the sole aim of enriching a bunch of yacht owners.\\-- Simon Stålenhag, quoted by Will Night, Wired \cite{knight_2022_2}
\end{quote}
On the contrary, some articles argue that taking inspiration from other artists' style is standard practice in art; intentionally or not, artists learn about and adopt other artists' styles and techniques \cite{harwell_2022, ibrahim_2022}. In a similar vein, others suggest that reordering and reusing content is a natural part of modern, digital creative practice \cite{lawson-tancred_2022}. One source argues that people are like visual sampling machines, thus operating similarly to \ac{TTIG} systems \cite{droitcour_2022}. Some artists oppose this by arguing that, even though copying others is common, \ac{TTIG} systems make it more evident and easy \cite{vox.com_2022}.

The second issue related to copyright concerns whether artists can take ownership of the artwork they create with \ac{TTIG} systems. Is it the developer of the algorithm who owns the image, or the artist who created the prompt? Currently, the art created with these models is not subject to copyright protection, meaning that the artist cannot claim rights for the generated image \cite{goldman_2022}. Matters related to copyright remain unclear, and lawsuits are underway \citep[e.g.][]{butterick_litigation}, but people argue that copyright laws will evolve with \ac{TTIG} and related systems, as happened previously with other technologies \cite{knight_2022_2}. One proposed solution is that artists could opt out of whether they would like to be in a database \cite{rajnerowicz_2022}.

\paragraph{Big Tech Critique} Some articles raise critique and expresss negative attitudes towards those technology companies who introduced \ac{TTIG} systems. Some artists express concern that 
it will be the companies \cite{knight_2022_2, gault_2022} who will profit from digital art creation in the future without compensating the artists whose work has been used \cite{knight_2022_2, gault_2022}:
\begin{quote}
    (...) it reveals that that kind of derivative, generated goo is what our new tech lords are hoping to feed us in their vision of the future. -- Simon Stålenhag, quoted by Will Night, Wired \cite{knight_2022_2}
\end{quote}
\begin{quote}
    Technology is increasingly deployed to make gig jobs and to make billionaires richer, and so much of it doesn't seem to benefit the public good enough. -- Matt Borrs, quoted by Matthew Gault, Vice \cite{gault_2022}
\end{quote}
\rev{Some companies providing \ac{TTIG} systems have responded to these concerns. In an interview, David Holz}, founder of Midjourney, states that in case of a notable dissatisfaction among the artists, it is possibly worth considering a payment structure to compensate those whose work is used in training the models \cite{claburn_2022}. 

\paragraph{Biased and Harmful content} Another issue raised by many concerns the inherited generation bias in \ac{TTIG} systems: given that the training data has been scraped from the internet, stereotypes are enforced \cite{macaulay_2022, rajnerowicz_2022, knight_2022, hertzmann_2022}:
\begin{quote}
    When I search, it defaults to white. It’s never given me a non-white person.\\-- Erin M Riley, quoted by Anna Furman, The Guardian \cite{furman_2022}
\end{quote}
Another perceived risk is that \ac{TTIG} systems can be exploited to produce harmful content in large quantities, including violent imagery, pornography, or deep fakes. At the time of conducting this survey, DALL·E 2 and Midjourney already implemented policies that seek to inhibit the generation of harmful content; the open-source system Stable Diffusion in contrast has been highlighted for not imposing such restrictions \cite{knight_2022, macaulay_2022, rajnerowicz_2022}.\\

In summary, our review of non-academic articles has revealed 11 topics: Creative process, Roles in the creative process, Using output as a source of inspiration, Prototyping/conceptualising ideas, Increased efficiency, Impact on creatives' jobs, Democratisation of creative work, Devaluation of creative work, Bias and harmful content, Copyright issues, and Big Tech critique. We use these topics as a priori codes and themes in our Thematic Analysis (Sec.~\ref{sec:study_ta}).\\

We note further shortcomings in this body of work, complementing our analysis of related academic work in Sec.~\ref{sec:related_academic}. While providing a broad view of the discussed issues, non-academic sources typically report only on the thoughts and feelings of a few professionals from various creative industries. The topic coverage remains shallow, often one-sided (e.g. purely optimistic, or very critical) and rarely conveys inner conflicts within individuals that most would deem natural. Moreover, these articles provide little insight into the underlying intentions of the publishing body; when, for instance, the authors of a blog post overlap (e.g.~in practice, or affiliation) with the purveyors of the system that is discussed, the question of bias arises. The interviewed professionals are often recruited selectively, and the corresponding recruitment strategies, including how the invitation was formulated, are not well documented. This lack of transparency attests existing non-academic related work little reliability to inform e.g.~policy development.

\section{Exploratory Interview Study}
\label{sec:study}

To provide a rich and deep account of all phenomena pertaining to professionals' perception, adoption and use of \ac{TTIG} systems, we chose to collect qualitative data from semi-structured interviews which was then further structured and made accessible through Template Analysis~\citep{brooks2015utility}, a specific form of Thematic Analysis~\citep{braun2006using}. 

Our choice of study epistemology positions puts the researcher into an active dialogue with the data. We thus transparently highlight individual contributions using the Contributor Roles Taxonomy \citep[CRediT,][]{credit}: Conceptualisation (4), Methodology (1,\rev{3,}4), Resources (incl.~recruitment; 1,2,3,4), Investigation (incl.~interviews; 1), Data Curation (incl. transcription and annotation; 1), Formal Analysis (1,4), Visualisation (1,4), Writing (1,2,3,4), Funding Acquisition (4), Supervision (4). Our study complies with the ethical principles of research with human participants laid out by the Finnish National Board on Research Integrity (TENK). We invested considerable effort in pseudonymising the interview data to the point where relevant information for the study would be lost; however, given the small size and transparent nature of Finnish game industry, a risk of the participant to be identified still remains. While we provide all remaining materials in the Supplementary, the raw and coded transcripts could thus not be made available.

\subsection{Recruitment}
\label{sec:study_recruitment}
We recruited our participants through (i) social media, and (ii) our personal and professional networks. We chose social media (i) to introduce our study to a maximally wide and diverse community throughout the country. More specifically, we advertised on two popular Facebook groups for digital artists and game industry professionals in Finland: \enquote{Finconauts - Finnish Digital Art Community} and \enquote{Play Finland}. To increase the diversity of our participant pool, we complemented this strategy by (ii) purposeful sampling \citep{palinkas2015purposeful} for roles and gender from our networks.

We carefully crafted our study invitations (\rev{Supplementary}) to optimise inclusivity w.r.t. three criteria. Firstly, we welcomed perspectives from different professional roles. Since Finnish game studios are often small in international comparison, professionals usually wear multiple hats and participate in more varied tasks than their job title might betray. At the same time, we wanted to ensure that our participants are actively involved in creative work and do not primarily comment by proxy. To accommodate both requirements, we omitted job titles and explicitly invited \enquote{diverse backgrounds and creative roles (including creative team leads)}. Secondly, we invited professionals with different levels of experience in engaging with \ac{TTIG} systems to mitigate potential bias from usage experience and AI literacy~\citep{long2020literacy,ng2021literacy}. Thus, we highlighted that \enquote{prior experience in using
these systems is not required}, and that any participant can get an introduction and opportunity to try \ac{TTIG} out first hand. Thirdly, we wanted to hear all views on the potential benefits and threats of these systems. At the time of our study invitation, debates on the use of creative AI and \ac{TTIG} systems specifically became very polarised. To welcome voices from all camps and any place in-between, we communicated with brief, neutral and inclusive language, and tried to hide the authors' individual positions in these debates.

We did not offer any study compensation in order to avoid bias and instead encourage people to participate primarily out of their interest to share their thoughts with us and our readers. Based on previous experience, we expected this mentality of sharing to be prevalent in Finnish game industry, which is confirmed by our successful recruitment.

\subsection{Procedure}
Our participants responded to the study invitation via email, and we fixed an interview time and place that was most convenient for them. Three days before the interview, we emailed our participants again with the request to review our privacy policy and fill in an online survey for providing informed consent and basic demographic data (age and \rev{identified} gender). Each participant was given an individual identification number to pseudonymise the survey and, later on, the interview data. We also inquired whether our participants had any prior experience in using \ac{TTIG} systems, and offered them a demo session with OpenAI's DALL·E 2. All refused, as every participant had at least some prior experience. Our templates for this email and the survey are provided in the \rev{Supplementary}.

In the interview sessions, we started with a casual chat that included a more detailed introduction to the study and the interview procedure. We asked for permission to record, before conducting the actual interview guided by our template (Sec.~\ref{sec:interview_materials}). All interviews were conducted during November and December 2022 by the first author. Nine sessions were conducted online, and five on-site. All sessions lasted for approximately one hour.

\subsection{Semi-Structured Interviews}
\label{sec:interview_materials}

We chose to conduct semi-structured interviews~\cite{adams2008questionnaires} to gather data on our overarching research questions (Sec.~\ref{sec:introduction}), while leaving room for participants to bring in the topics that they care most deeply about. Our interview outline \rev{(Appx.~\ref{sec:appx_interview})} was developed between all authors, involved the help of game design and game industry practice researchers, and was informed by 
the User Acceptance of Information Technology model (UTAUT) \cite{venkatesh2003user}. 
In particular, we formulated a subset of questions to overlap with the UTAUT questionnaire items \enquote{Performance / Effort Expectancy}, \enquote{Attitude Toward Using Technology}, \enquote{Social Influence}, \enquote{Facilitating Conditions}, and \enquote{Behavioral Intention to Use the System}. The interview consisted of three parts: 
\begin{enumerate}
    \item \textbf{Warm-up}: Questions related to education, industry experience, previous and current roles, and tasks. We asked all participants \enquote{What comes to your mind when you hear the term AI image generator?}, and followed up with questions on the specific systems used, usage frequency and context of use. To better understand our participants' level of AI literacy, we also asked about their experience in using other AI systems.
    \item \textbf{Main part}: Structured based on our overarching research questions (Sec.~\ref{sec:introduction}), we inquired about our participants' (i) perception, as well as (ii) current and (iii) future use of \ac{TTIG} systems. Amongst others, we asked about the value and concerns associated with these systems (i); how our participants employ them in their individual creative processes, and to what end (ii); and how they expect this to change in the future, including pressures w.r.t. learning about and using the systems, and effects on roles and outsourcing practices (iii).
    \item \textbf{Wrap-up}: To indicate the end of the interview and distil what our participants felt most strongly about, we asked \enquote{If you could pick one thing people should know about these systems, what would it be?}. Finally, they were offered to further comment on points which, in their opinion, have not received sufficient attention.
\end{enumerate}
Following a pilot study with one participant, we refined the template to 19 primary questions (\textbf{bold} in the template). These were asked in different order to guide the conversation, while allowing for a natural flow of discussion. 

\subsection{Template Analysis}
\label{sec:study_ta}

We automatically transcribed our interview recordings using Office 365, followed by a manual editing pass to correct transcription errors, annotate affect and nonverbal cues, and anonymised the data by redacting personal details. 
Two of the authors then conducted a Template Analysis \citep{brooks2015utility} on the data, adopting the six steps defined by \citet{brooks2015utility} to enable asynchronous, collaborative coding. A specific form of Thematic Analysis \citep{braun2006using}, Template Analysis supports the creation of the eponymous coding template (Supplementary), i.e., a hierarchy of codes and themes, from the start and based on a subset of the data. The initial coding template can comprise a priori codes, and is refined throughout the coding process. This enabled us to include the topics identified in Sec.~\ref{sec:related_work} and thus connect our findings to the existing, albeit more shallow, discussion in related work. Template Analysis moreover allows for deeper code hierarchies, enabling us to capture richer and more detailed aspects of our data. Our epistemological stance is \enquote{subtle realist}~\citep{brooks2015utility}, and we coded most of the data inductively to explore previously unknown phenomena. We provide further details on our analysis and describe the exact coding procedure in Appx.~\ref{sec:appx_analysis}.

\begin{table}[b!]
    \centering
    \caption{Participant overview and categorisation of system usage based on \link{US.2:} \textit{Usage frequency}. Note that this is a rough and incomplete categorisation based on qualitative statements. No usage was specified by participant 9, despite being asked.}
    \begin{tabularx}{\textwidth}{llXl}
        \toprule
        Id & Demographics & Usage Statement & Usage\\
        \midrule
        1 & M, 34-44, VFX Lead & I think I like almost daily. & +++\\
        2 & M, 45-54, Art Director & I'd say it's daily. & +++\\
        3 & F, 34-44, UX Designer & Yeah, it's it's up and down the use. Haven't really played with it that much as much others have used it a lot more. & +\\
        4 & M, 34-44, Game Designer & I don't use it everyday (...). Maybe a little more than once a month. & +\\
        5 & M, 34-44, Artist & Well, basically I use them actually quite often. & +++\\
        6 & M, 25-34, UX Designer & I would say monthly. I use them in bursts so it's usually like one or two days a month I use them a lot and then the rest I don't really. & ++\\
        7 & F, 25-34, Senior Game Artist & So do you use these systems like frequently or regularly? No, I don't. & +\\
        8 & M, 34-44, Senior Art Director & I use those daily. & +++\\
        9 & F, 34-44, CEO/Creative Director & No data available. & $\otimes$\\
        10 & M, 34-44, Art Director & Yeah, I use almost daily. & +++\\
        11 & M, 34-44, Narrative Designer & Yeah, I'd say, uh, I'd say a couple of times a week. & ++\\
        12 & M, 45-54, Lead Artist & It's it's weekly. More like weekly. & ++\\
        13 & M, 34-44, Concept Artist &  I don't know on and off for a month. & ++\\
        14 & M, 34-44, Lead UI/UX Designer & Once a week, once a month. Three times a week. It depends, really. & ++\\
        \bottomrule
    \end{tabularx}
    \label{tbl:participants_overview}
\end{table}

\subsection{Participants}
\label{sec:study_participants}
14 participants took part in our final study, of which half were recruited through social media and the other half through our networks. In our demographics questionnaire, 11 participants identified as male (79\%), and three as female (21\%). While being highly skewed, this gender ratio is representative of the Finnish game industry \rev{\cite[22\% identifying as female in 2020/22,][]{hiltunen_2020, hiltunen_2022}}. Three of our participants were aged between 25-34, nine between 34-44, and two between 45-54. 

We learned more about our participants' industry experience through our interviews. Everyone told us about their time spent in the industry: at least 1.5 years, at most 24 years, and 12 years in the median. 
Their current roles included Generalist/Game/Concept Artist (4), Art Director (3), UX Designer (3), Game Designer (1), Narrative Designer (1), VFX (1), and CEO / Creative Director (1). Three assumed lead positions. They came from 12 different game studios across Finland, with size varying from small/indie (8, <100 employees), over mid-sized (3, 100-200 employees), to big studios (3, > 200 employees). The studios covered a wide range of platforms and genres, but this data is incomplete. People developed for mobile (5), PC/console (3), PC only (1), and all of these platforms (5). Game genres included Massive Multiplayer Online (3), Puzzle (3), strategy (1), Action (1) and Card (1) games, as well as combinations of those (5).

\begin{figure}[htbp]
\includegraphics[width=\linewidth]{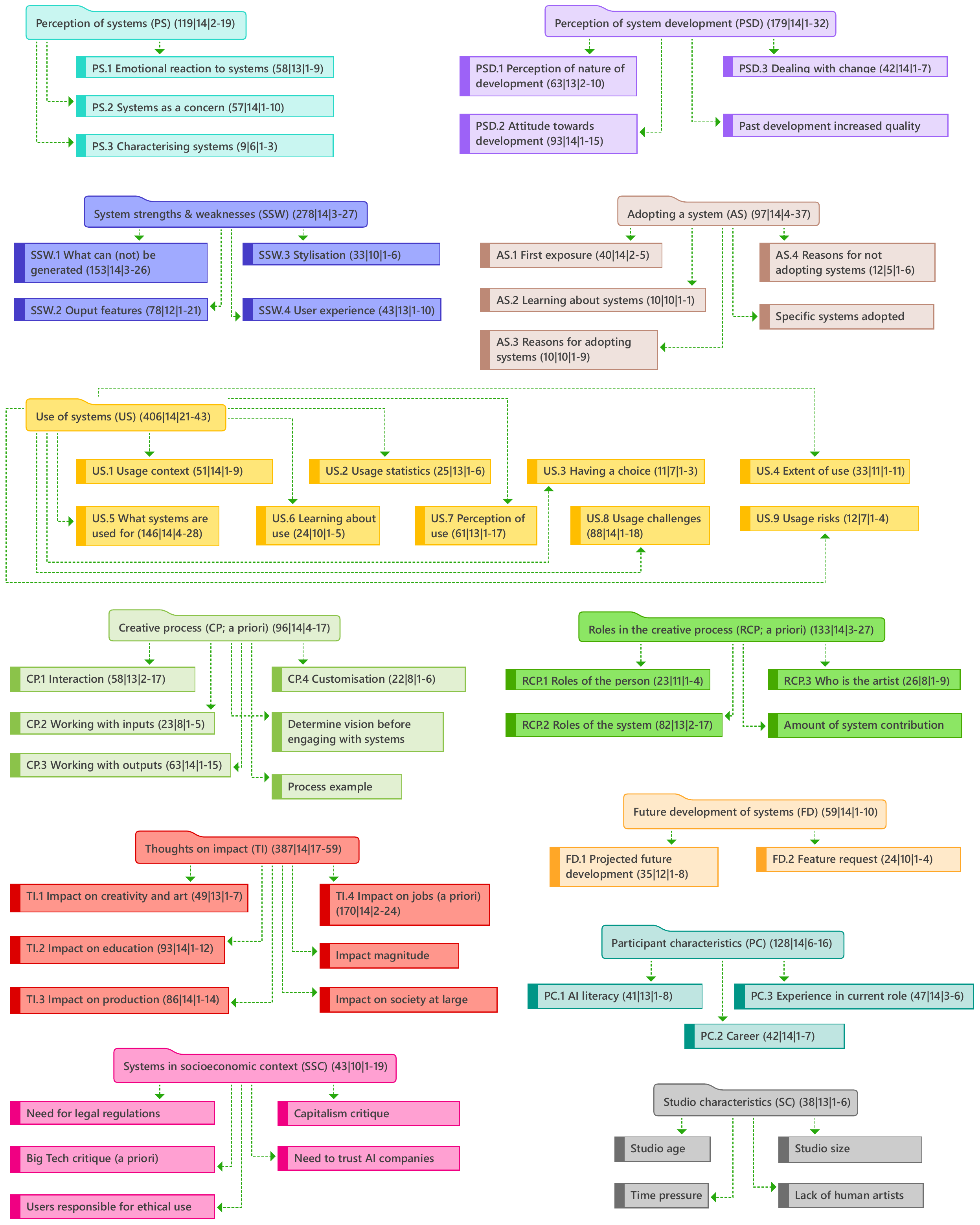}
\caption{\rev{Thematic map of all 12 identified themes with their grounding, and one additional level of hierarchy.}}
\label{fig:themes_overview}
\end{figure}

\rev{As part of our semi-structured interview, we also inquired about our participants' extent of \ac{TTIG} system usage. Tbl.~\ref{tbl:participants_overview} summarises the demographics along with a categorisation of usage frequency based on evaluating the respective statements. By complementing illustrating quotes in our results with the participant Id, Demographics and Usage data, we support the reader in evaluating statements against a participants' individual backgrounds.}

\section{Results}
\label{sec:results}

Through our Template Analysis, we identified more than 300 codes, structured into 12 themes, 39 sub-themes, and further underlying themes (up to four levels of hierarchy). Each interview has been coded between 110 and 233 times ($\mu=176.3, \sigma=37.5$). We provide an overview of all top-level themes \rev{and one additional level of hierarchy in Fig.~\ref{fig:themes_overview}. Sub-themes are enumerated, while codes are not. The formatting of the boxes is a software artefact and can be ignored.} Here, we also specify the grounding for each (sub-) theme the form (total grounding over all interviews | in how many interviews | min - max grounding across interviews). We summarise all top-level themes, including their short description and the research questions they contribute to, in Tbl.~\ref{tbl:themes_overview}. 
It is indicative of a successful, primarily inductive Thematic Analysis that the identified themes do not simply correspond to the researchers' questions~\citep{braun2006using}. This also applies here. 
In all descriptions, \enquote{system} defaults to \ac{TTIG} systems. \rev{We provide the full coding template with a detailed description of all themes in the Supplementary.} \rev{Illustrative quotes have been edited only slightly: for brevity, we omitted parts that do not relate to the respective theme (indicated by \enquote{(...)}); to improve readability, we removed repetitions and utterances that do not clearly convey a specific sentiment, or add to the interpreted meaning.}


\rev{We have not collected data on the \ac{TTIG} system versions that our participants had used up to their interviews (November-December 2022). To better understand the systems' capabilities when referred to as \enquote{current} or \enquote{previous}, we document the version history of the most popular systems up to, and including, the interview period in Tbl.~\ref{tbl:ttig_versions}.}

Our participants considered \ac{TTIG} \enquote{the biggest thing to happen for human creativity} (3, F, 34-44, UX Designer, +), \enquote{changing the world as we speak} (3, F, 34-44, UX Designer, +), and already revolutionising the way people work (8, M, 34-44, Senior Art Director, +++). While some do not see an immediate effect on their work, they yet project a \enquote{gigantic} impact on game industry as a whole (6, M, 25-34, UX Designer, ++) (\link{TI}: \textit{impact magnitude}). We shed light on the nature of this development by summarising the contribution of our themes to each of our research questions (Sec.~\ref{sec:introduction}). We conclude this discussion by comparing our findings to those in related work (Sec.~\ref{sec:related_work}). More details on each code (in \textit{italics}, inflected), and the corresponding themes (in \textbf{bold}) can be found in the full coding template \rev{(Supplementary)}. Only a detailed description of the identified themes brings a Thematic Analysis to life;  we hence strongly recommend our readers to use the coding template as a reference when reading the following summary.


\FloatBarrier
\begin{table}[htbp]
    \centering
    \caption{Themes identified through our Template Analysis, together with the research questions (RQ) they contribute to, and a description of the theme. \rev{We provide the entire codebook with sub-themes down to the code level in the Supplementary.}}
    \begin{tabularx}{\textwidth}{llX}
        \toprule
        Theme & RQ & Description\\
        \midrule
        Perception of systems (PS) & 1 & Participants' perception of, and reactions to TTIG systems in their current form.\\
        Perception of system development (PSD) & 1 & Participants’ perceptions of, and attitudes towards the past and future development of TTIG systems.\\
        System strengths \& weaknesses (SSW) & 1,2,3 & Perceived strengths and weaknesses of TTIG systems w.r.t. the overall system, its outputs, and the user experience.\\
        Adopting a system (AS) & 2 & Factors in participants' adoption of TTIG systems for their creative practice, including the how and why of discovering, adopting and staying with a specific system.\\
        Use of systems (US) & 2 & Participants' current and projected future use of TTIG systems.\\
        Creative process (CP; a priori) & 2,3 & Roles that our participants assume for themselves and the system in the creative process.\\
        Roles in the creative process (RCP; a priori) & 1,2 & Roles that our participants assume for themselves and the system in the creative process.\\
        Thoughts on impact (TI) & 1,3 & Participants' thoughts on the form and nature of TTIG systems' current and future impact.\\
        Future development of systems (FD) & 3 & Participants' thoughts and wishes toward TTIG's future.\\
        Participant characteristics (PC) & 1,2,3 & Indicators of individual participants' level of AI literacy, as well as information on their educational background, career, and experience in their current role.\\
        Studio characteristics (SC) & 1,2,3 & Insights characterising our participants' current work reality.\\
        Systems in socioeconomic context (SSC) & 1,2,3 & Participants' views on TTIG in society and economy at large.\\
        \bottomrule
    \end{tabularx}
    \label{tbl:themes_overview}
\end{table}


\subsection{What Are Professionals’ Perceptions and Attitudes Towards \ac{TTIG} Systems and Their Future? (RQ1)}
\label{sec:discussion_rq1}

\subsubsection{Perception of Systems}
The majority of our participants stated to be \textit{impressed} by, or \textit{interested} in \ac{TTIG} systems, and several showed \textit{excitement} (\link{PS.1}):
\begin{quote}
    And I was super excited about this. Like this, kind of being able to console thousands of artists with one prompt and getting these fresh ideas and like getting ideas that you'd never come up with yourself. -- (1, M, 34-44, VFX Lead, +++)
\end{quote}
These mostly positive, first-hand reactions contrast with the majority of our participants expressing \textit{concerns about people losing their job or livelihood} (\link{PS.2}). While only some admitted to be scared, the majority of our participants expected or knew \textit{others to be scared} (\link{PS.1}): 
\begin{quote}
    The first impression probably was the same as every single artist that it can't do... You know you can't do stuff with the AI that I can do? It's just stealing from other artists and... Fuck, I'm gonna lose my job. Every single artist I've been talking to about this has basically had the same reaction at first. -- (5, M, 34-44, Artist, +++)
\end{quote}
In a comparatively weaker theme, few participants have been anthropomorphising the systems, either to downplay them as \textit{dumb}, \textit{emotionless} and \textit{soulless}, or characterising them as having \textit{opinions and personality} (\textbf{PS.3}). More than half raised \textit{ethical concerns}, primarily concerning copyright issues (\link{PS.2}). A major challenge in using \ac{TTIG} systems today, many of our participants expressed unease in using a system which leverages their fellow artists' work as \textit{training data} without adequate compensation. Almost as many feel inhibited by the question of \ac{TTIG} \textit{output ownership and usage} in a commercial context being under dispute. At the same time, companies' legal departments, as well as public legislation, are perceived as \textit{adjusting too slowly} (\textbf{US.8.1}). While only few participants deny that \ac{TTIG} systems are \textit{worth their cost}, mostly based on a lack of need, more are affirmative, with one highlighting the systems' value particularly for independent artists. Others expressed reluctance to pay if their contribution would not benefit the artists whose work was used for model training (\link{US.7}), thus indicating \textit{conditional use} (\link{US.4}) based on \textit{ethical concerns} (\link{PS.2}).  

\paragraph{Perception of \ac{TTIG} development}
Our participants perceive the development of \ac{TTIG} systems (Sec.~\ref{sec:background}), in particular its \textit{increasing capabilities and output quality}, not only as \textit{fast}, but also with \textit{uncertainty}, feeling overwhelmed (\link{PSD.1}):
\begin{quote}
    I feel like stuff that was impressive even a month ago or two months ago feels like really crappy by today's standards. It's that the the speed is so rapid, so fast.\\-- (8, M, 34-44, Senior Art Director, +++)
\end{quote} 
Few participants emphasised that the \textit{development cannot be stopped} (\link{PSD.1}), saying that \enquote{the box is now open} (12, M, 45-54, Lead Artist, ++) and that they \enquote{don't see any way going back} (8, M, 34-44, Senior Art Director, +++). The majority shared \textit{mixed feelings} about the development of \ac{TTIG} systems; most combined a \textit{pragmatic}, mixed with either a \textit{postive and optimistic}, or a \textit{critical} attitude (\link{PSD.2}). We find this positive, or at least pragmatic, attitude to be rooted in an industry culture of \textit{embracing progress} (\link{PSD.3}); while advocates of \ac{TTIG} more generally refrain to historical examples of technological disruption (Sec.~\ref{sec:related_work}), several of our participants drew a much closer parallel to the industry's long tradition and \textit{continuity in the use of tools and \ac{AI}}. Notably, multiple participants seem to prioritise \ac{TTIG}'s benefits for the final game over the technology's potential impact on their role and creative practice (\link{PSD.2}. Only one person was downright, albeit somewhat reluctantly, \textit{opposing progress} (\link{PSD.3}).

\subsubsection{Perceived Role in the Creative Process}
Most of our participants agree that their role, when interacting with \ac{TTIG}, transforms fundamentally, e.g.~from a first-hand craftsperson to a \textit{director to the AI}. Orthogonal to the role as director, one artist criticises the shift in tasks toward editing an AI's output as demotion to the role of \textit{assistant to the \ac{AI}} (\link{RCP.1}):
\begin{quote}
    If I were using something like Midjourney, the latest version, which is quite good, I would probably get an image of a cyberpunk guy that we could just use as is as long as you fix some of the errors like OK, the jacket is the wrong way on this side. And OK, this this hand is too many fingers and the the hair here is wrong, so I'd be fixing these little things and, so I would be just sort of the the assistant to the AI (...) The further one gets to the heart of the concept, having like an idea the more heartbreaking it is to hand it over to an AI. So to be just, you know, the publisher. Somebody who sort of fixes the errors is not a very appealing idea. -- (11, M, 34-44, Narrative Designer, ++)
\end{quote}
Almost everyone considered \ac{TTIG} yet another \textit{tool} in their creative process, rather than a creative partner in their own right. The majority sees these tools as \textit{allies} in their work, and more than half even considers them \textit{enablers}, capable of taking their creative work to new lengths. Crucially though, several of our participants also consider them potential \textit{competitors} for their jobs (\link{RCP.2}). Our participants' perception of prompt engineers, a potentially new role in games production, is divided: some see the role as tedious and uninspiring, some discard its artistic value altogether, and others draw a parallel to conceptual (rather than concept) artists (\link{RCP.3}). Closely related, we witnessed different views on the use of \ac{TTIG} systems being \text{fun} and \textit{rewarding}. Some participants emphasised the work with \ac{TTIG} to be fun, and to alleviate them of tedious, unrewarding tasks, putting emphasis on the final product. Others though prioritise the process in their creative work, and argue that \ac{TTIG} takes away what they deem most rewarding (\link{US.7}). Here, the following view of a team lead seems in stark contrast with the artists' view expressed in the previous quote:
 \begin{quote}
    We definitely can use just straight out stuff from Midjourney and give them to an artist instead and they'll be super happy now. This is a good idea and they can, like copy, paste and basically paint over stuff then maturing, and then we can use that already.\\-- (14, M, 34-44, Lead UI/UX Designer, ++)
\end{quote}

\subsubsection{Perceived System Strengths and Weaknesses}
Our participants were unanimous in considering the \textit{creation of more content} in less time a strength of these systems. However, they expressed opposed views on other capabilities, e.g. whether \ac{TTIG} is good at creating specific things beyond the capabilities of image search (\link{SSW.1}). We doubt that this disagreement stem from different quality standards, rather than people's varying flexibility toward realising their exact vision, the task and intended use of the output (e.g. concept vs. production-ready art) (\link{US.5}), and individual prompt engineering skills (\link{TI.2}), amongst others. Contradictory, most participants lamented that \ac{TTIG} cannot provide \textit{production-ready content} (\link{SSW.1}), while at the same time projecting a substantial impact on production (\link{TI.3}). Related, our participants agree that the systems' outputs are still very \textit{random}; they consider randomness both a strength and a weakness (\link{SSW.2}), in that it can support \textit{inspiration} (\link{US.5}) and \textit{exploration} (\link{CP.1}), but might also hold up progress in achieving a desired outcome. More than half of our participants discussed which styles \ac{TTIG} can or cannot produce well, e.g. \textit{fantasy} vs. \textit{realistic} styles. Several participants noted problems with the systems supporting stylistic \textit{consistency} across outputs, e.g. to produce assets consistently in the defined style of the game (\link{SSW.3}). While the systems can inform a particular style for a project, half of our participants noted issues with producing a \textit{specific} and uncommon style which might have been defined for the project ahead of time (\link{SSW.3}). Commenting on strengths and benefits of UIs and UX, several people support the current systems'  \textit{ease of use}; however, some also see opportunities for improving interfaces to support \textit{user-friendliness for artists} specifically. In particular, artists seem to struggle working with text, rather than a more familiar modality such as visuals or spoken words (\link{SSW.4}).

\subsection{How Are \ac{TTIG} Systems Adopted and Used in the Creative Practice Now? (RQ 2)}
\label{sec:discussion_rq2}

\subsubsection{System Adoption}
Most of our participants were \textit{first exposed} to \ac{TTIG} between Spring and Autumn 2022, i.e. \enquote{when it became a thing} (7, F, 25-34, Senior Game Artist, +), and thus cannot be considered early adopters. A few though, mostly artists, have been experimenting with much older \ac{TTIG} and, more generally, creative AI systems, for up to five years. \rev{Within the more common adoption timeframe,} most started experimenting with systems such as DALL·E 2, Midjourney, and Stable Diffusion (\link{AS.1}, cf.~Sec.~\ref{sec:background}). Since most participants only wanted to try out \ac{TTIG} first hand, some were initially \textit{indifferent} w.r.t.~which specific system to take on, and chose by \textit{availability}, which was limited by wait lists and high server loads. Almost half of our participants indicated to having adopted a specific system for requiring \textit{no installation}, and one participant because it used a \textit{familiar interface} such as Discord (\link{AS.3}).

\subsubsection{Context and Extent of Use}
Only about a third of our participants use \ac{TTIG} on a daily basis, another third a few times per week, and another third only monthly (Tbl.~\ref{tbl:participants_overview}). Given this \textit{usage frequency}, not all statements should be understood to arise from a pattern of regular, intense use (\link{US.2}). The majority of our participants only explore the systems for future \textit{work use}, rather than employing them actively in production. Almost everyone mentioned \textit{free time} use, also because they find their use-cases at work too limiting (\link{US.1}). From those commenting on this issue, the majority understood the present use of \ac{TTIG} as \textit{optional} and not mission-critical, but several expect the use of \ac{TTIG} systems to become \textit{mandatory} in the future, dictated by market forces (\link{US.3}). Several of our participants appeared \textit{dedicated} to using \ac{TTIG} now or in the future of their work. Crucially though, the majority is still exploring and \textit{pondering} its use (\link{US.4}). Few participants did \textit{not feel a need} to adopt the system, e.g. because their current \textit{project was already too advanced} to bring in new and often inconsistent technology (\link{AS.3}). One Senior Artist hesitated in adopting a system because they were unsure about its work use being eligible, perceiving it as \textit{cheating} (\link{US.7.2}). Some participants refused to use \ac{TTIG} systems in their \textit{free time} (\link{US.1}) because they consider it unrewarding (\link{US.7.2}). One artist \textit{rejected} the systems altogether (\link{US.1}) because they perceived them as unethical (Sec.~\ref{sec:discussion_rq1}):
\begin{quote}
    Yeah, then [when the second generation of systems came about] I started to feel a mild repulsion [laughs]. Like, OK, this is now taking work away from people who have been trying to create their craft for even 20 years. So that was something that I really, I hit a wall with I really didn't want to do any AI art after that (...).\\-- (12, M, 45-54, Lead Artist, ++)
\end{quote}
Moreover, two participants described their \textit{conditional} use of \ac{TTIG} up to a certain point, motivated by ethical concerns. Similarly, more than half of our participants decided, motivated by legal and ethical concerns, to use a system's output only \textit{separately from the final product} (\link{CP.3}):
\begin{quote}
    I made sort of like a responsible decision for myself, limit using AI only to get a starting point or to get a new direction to something that I cannot get forward with so (...) I never use [TTIG to produce a final image] because that in my opinion crosses the line of taking work or the property from real people to the public domain. In a way that the people who have been trying to craft their work for years are not getting paid.\\-- (12, M, 45-54, Lead Artist, ++)
\end{quote}

\subsubsection{What Systems Are Used For}
Almost everyone emphasised using the systems' outputs as a \textit{source of inspiration}, and to \textit{prototype / conceptualise ideas}. The latter has especially been popular with professionals in a lead role to \textit{pitch} their ideas to clients or their team, thus \textit{supporting communication} of what would otherwise be hard to put into words (\link{US.5}). Some participants mention that they \textit{sidestepped their artists} in this process, e.g. because they are too few, too busy, or because engaging with them is considered too time consuming. One participant chose to adopt a \ac{TTIG} system in the first place because \textit{artists were unavailable} (\link{AS.3}). Next to using \ac{TTIG} for inspiration and art concepts, more than half of our participants indicate its use to \textit{create production-ready art} (\link{US.5}).

\subsubsection{Creative Process} 
The majority of our participants state to interact with \ac{TTIG} systems by \textit{iterating prompts and outputs} -- arguably the most common practice across creative domains. Crucially though, they instantiate it in two ways: given a specific task, people approach the interaction with the system systematically and stress the need to \textit{determine a clear vision} beforehand (\link{CP.1}). However, if seeking \textit{inspiration} or \textit{serendipity} (\link{US.5}), they engage in a more \textit{exploratory and playful} manner (\link{CP.1}):
\begin{quote}
    It's more like play where you... ride the wave of the AI and see where it takes you.\\--  (1, M, 34-44, VFX Lead, +++)
\end{quote}
Partly due to \ac{TTIG}'s inherent randomness (\link{SSW.2}), half of our participants said to interact with the systems not with a clear plan, but through \textit{trial and error}. Several professionals also emphasised the value of \textit{generating multiple variations, and then filtering} for the best version. This is typically used to ideate and kick-off the artist's own creative process, or that of their team (\link{CP.1}.
While \textit{using artists' names as prompts} still seems a common practice with more than half of our participants (\link{CP.2}), few individuals mitigate their \textit{ethical concerns} (\link{PS.2}) by instead \textit{deconstructing and extending a given style} (\link{PS.2}), a process which is arguably closer to established practices in the art world (cf.~\ref{sec:related_work}):
\begin{quote}
    (...) I do think that it's unethical to use somebody's name as part of your prompt, but I don't see any ethical problems to ending up with images that look like somebody else's. If you've done that kind of deconstruction, and you're prompting in a way that details, that you want these kind of brush strokes. You want this kind of lighting. You want this kind of stylisation. You want these kind of things and you want to include this and that. Then I think prompt engineering becomes the creative process.\\ -- (8, M, 34-44, Senior Art Director, +++)
\end{quote}
On the opposite end of fine-tuning, and related to playful exploration (\link{CP.1}), few participants stated to \textit{provide under-specified prompts to provoke surprise} (\link{CP.2}). Almost all of our participants expressed the need to \textit{further edit a system's outputs} (\link{CP.3}), due to flaws in quality (\link{SSW.1.2/2.2/3.2}) or a mismatch with their artistic intentions. While the human artist is thus still under high demand, their involvement in creating one specific artwork is drastically reduced; two lead artists (14, M, 34-44, Lead UI/UX Designer, ++); 2, M, 45-54, Art Director, +++) estimated the AI's work to account for 80-90\% of the final piece. The majority of our participants however consider an artist's \textit{background and education} still relevant when using \ac{TTIG} in the creative process. For instance, artistic intention is required to guide the interaction, prompt engineering benefits from knowledge of art history styles and artistic techniques, and assessing the system's output requires an intuitive understanding of aesthetics and rules to achieve e.g. colour harmony (\link{TI.2}). Also related to education, our participants have raised a \textit{lack of technical expertise} as another major challenge in employing \ac{TTIG} systems in the creative process (\link{US.8.1}). This affects both their freedom in experimenting with systems that need more sophisticated setups, and their ambitions for customising \ac{TTIG} in-house (Sec.~\ref{sec:discussion_rq3}).

\subsection{How Will \ac{TTIG} Systems Change and Be Used in Future Creative Practice? (RQ 3)}
\label{sec:discussion_rq3}

\subsubsection{Impact on Art \& Creativity}
The majority of our participants expressed excitement about \ac{TTIG}'s potential to \textit{democratise creative work}, notably not only to enable people who previously lacked the necessary skills, but also those who lost them, or do not have the means to develop them. At the same time, some also showed concern about a \textit{devaluation} of professional skills and the value of their work. While the appreciation of skill might diminish, several of our participants expect \textit{ideas to be valued more} (\link{TI.1}).

\subsubsection{Impact on Education}
Few express concern that this devaluation of creative work could \textit{discourage people from learning and applying artistic skills}. This however only holds for traditional skills; despite not considering the use of \ac{TTIG} systems as mandatory yet (cf. Sec.~\ref{sec:discussion_rq1}), most of our professionals stress that \textit{artists need to learn and adapt to \ac{TTIG} systems}. To facilitate this transition, our participants call for \textit{educational institutions to adapt} \link{TI.2}:
\begin{quote}
    Oh yeah, I mean. It's gonna be a huge added asset to the tool set and I think everybody who's studying today should keep their eyes on what's going on. (...) and unless all of the educational institutions are ready for this (...) change, they'll be basically educating people who are not going to be able to find jobs. -- (2, M, 45-54, Art Director, +++)
\end{quote}

\subsubsection{Impact on Production}
Most prevalent amongst the related codes, people expect \ac{TTIG} to yield \textit{more} art in less time (i.e., more \textit{efficiently}). This applies to communicating ideas and generating concept art, but also to creating assets for production. Crucially, some argue that this does not necessary have to imply job losses, but could also allow for \textit{more games} to be produced, with \textit{higher quality}, or of entirely \textit{new type} (\link{TI.3}):
\begin{quote}
    There will be more visual content and it will be of higher quality and, uh, I think we will see more smaller teams able to do much larger sort of higher production quality and larger experiences or content. -- (2, M, 45-54, Art Director, +++)
\end{quote}
While half of our participants believe that production \textit{costs will decrease}, few argue for production \textit{costs to increase} because either high-end content, e.g. in AAA productions, cannot be produced by existing models, or because using \ac{TTIG} will require more iterations at the same price (\link{TI.3}).

\subsubsection{Impact on Jobs}
None of our participants doubted that \ac{TTIG} will lead to substantial changes to people's jobs, first and foremost a \textit{transformation of roles} (\link{RCP}). From all specific roles (\textbf{TI.4.3}), people expect the technology to \textit{impact illustrators' jobs} the most. However, our interviewed professionals appear to disagree about whether these systems will \textit{create new jobs} or \textit{replace} them. While those in a position to make actual \textit{hiring decisions} agree that less artists could he hired, some seem to rule this option out for ethical reasons (\link{TI.4}).
\begin{quote}
    Well, I guess it affects that instead of hiring five concept artists now there's need for only three or two or something like that. So, it might might have that kind of effects on the industry. -- (10, M, 34-44, Art Director, +++)
\end{quote}
\begin{quote}
    Like lot of questions there when it comes to living artists. First of all, living artists need definitely support, meaning the sort of actual financial support to live as a human being. So I would always pay human over any machine.\\-- (9, F, 34-44, CEO/Creative Director, $\otimes$)
\end{quote}
In this context, almost half of our participants emphasise the potential for \ac{TTIG} to \textit{free time for more fulfilling work} (\link{TI.4}). Crucially though, our earlier findings (Sec.~\ref{sec:discussion_rq1}) suggest that, what people consider the most fulfilling part of their creative process, differs much between individuals.

Our data supports that \ac{TTIG} systems will change how professionals \textit{collaborate} in an intricate way (\link{TI.4.2}). One the one hand, our participants expect \ac{TTIG} to make some communication with artists unnecessary, in that now a team lead can create art for their pitch quickly themselves; on the other hand, this promises to enable a more effective communication of ideas between leads and their teams, or between colleagues.

Crucially, all of our participants emphasised that \textit{human artists are still needed} (cf.~Sec.~\ref{sec:discussion_rq2}). They can \textit{communicate and take feedback more naturally} and \textit{deliver exactly what is needed}; moreover only people at present can \textit{take the systems' output further and make it perfect}, introducing \textit{novelty and uniqueness}, and \textit{fitting the created art into a game as a whole}. Most notably though, many emphasised the value of their colleagues in offering a \textit{human connection} (\link{TI.4.3}):
\begin{quote}
    We are humans. We want to communicate others how we feel, what ideas we have and that's what it is. It's an accelerator for human ideation and being able to reach.\\-- (3, F, 34-44, UX Designer, +)
\end{quote}

\subsubsection{Future Development} Our participants expect future generations of \ac{TTIG} systems to overcome many of its current weaknesses. Moreover, they expect them to be \textit{integrated into existing tools} (\link{FD.1}). Our participants articulated the wish to customise \ac{TTIG} to fit their needs for e.g.~delivering a unique style (\link{US.9:} \textit{Risk of sameness}), or for including their own brands and IP (\link{SSW.1.2}) in the art. However, the proposed means, namely \textit{training their own AI} and \textit{using \ac{TTIG} in a pipeline with other AI} require technical competencies in teams which, as of yet, seem mostly absent (\link{US.8}).

Mirroring their expectations toward the systems' future development (\textbf{FD.1}), our participants' feature requests focus on better UIs and UX, including more \textit{natural UIs}, more \textit{UI elements affording fine control}, and more \textit{instant feedback} (\link{FD.2}). Together, these features would move AI closer to how we would expect to interact with a human artist. If this were to happen, we would expect it to become more difficult, albeit not impossible, for leads to hold on to human artists.

\section{Comparison to Related Work}
\label{sec:related_work_comparison}

\rev{We recovered and further substantiated all a priori themes from non-academic related work (Sec.~\ref{sec:related_non_academic}). Here, we compare our findings to those in previous and concurrent academic work (Sec.~\ref{sec:related_academic}).} 

We confirm four of the potentialities identified by \citet{ko2022large}, the use of \ac{TTIG} as an (1) image reference search tool (\link{SSW1.1:} \textit{Specific things [that cannot be searched for]}), (2) enabler of visual communication (\link{US.5:} \textit{To support communication}, (3) inspiration engine (\enquote{Rectifying Human’s Biased Creation}; \link{US.5:} \textit{As a source of inspiration (a priori)}) and (4) means for novice artist (\link{TI.2:} \textit{Systems compensate lack of skills}) to prototype their work (\link{US.5:} \textit{To prototype / conceptualise ideas (a priori)}). We can moreover support one of the identified weaknesses, namely that \ac{TTIG} systems (4) are inefficient and become a burden, albeit this impression is only shared by few of our participants (\link{US.7.2:} \textit{Perceiving systems use as not fun / not rewarding / not worth the effort}). We cannot confirm the limitation that \enquote{LTGMs only generate predictable images}; in fact, most of our participants emphasised the randomness of their outputs and lack of controllability (\textbf{SSW.2.1/2}: \textit{Randomness}), both as a strength (e.g. for inspiration) and as a weakness (for realising a specific creative intention). Moreover, we provide a more nuanced assessment of the statements that \ac{TTIG} systems (2) do not support personalisation, and (3) restrain creativity through the prompting mechanism. While personalising \ac{TTIG} systems is possible and demanded by our participants (\link{CP.4:} \textit{Training your own AI's}), it is currently limited by the expertise in the teams (\link{US.8:} \textit{Lack of technical expertise}). While text prompting appears constraining to some of our participants who consequently demand more natural user interaction (\link{FD.2:} \textit{More natural UI}), many also highlight \ac{TTIG}'s existing ability to enhance creativity (\link{RCP.2:} \textit{Systems as enablers / superpower}).

\rev{Our comparison to \citet{inie2023} should be taken with a grain of salt due to their missing focus on \ac{TTIG} systems, amongst others (cf.~\ref{sec:related_academic}). We can confirm that the advancement of creative \ac{AI} \enquote{prompts important reflections on what defines creativity and how creatives imagine using AI to support their workflows} (\link{CP/TI}). We particularly note parallels in participants accepting that AI should be considered creative, while also expressing resentment. Moreover, our participants similarly appeared worried to very different degrees. We can support two of the identified reasons through our data. Their participants similarly highlight (1) weaknesses in output quality (\link{SSW.1.2/2.2/3.2}) to motivate why people are still needed in the process (\link{TI.4.3:}\textit{To take it further \& make it perfect}). Interestingly, their participants lament (3) copyright issues primarily w.r.t.~the training data, but do not express the need for obtaining copyright on the generated outputs (\link{US.8.2:} \textit{Copyright issues with output ownership and usage}). This serves as one example of how our focus on a specific industry can generate richer, actionable insights. Considering reasons to not worry about creative AI, we similarly observed that our participants highlighted the (1) reliance of \ac{TTIG} on human input and, in the same context, emphasised the need for original thinking (\link{TI.4.3:} \textit{To provide novelty \& uniqueness}). We can only provide limited support to their sub-theme that (2) AI output is not convincing, given our participants' mixed statements on \ac{TTIG}'s shortcomings (e.g.~\link{SSW.1.2:} \textit{(Lacking) true novelty}; \link{SSW.2.1:} \textit{Not looking man-made}) while also emphasising production use (\link{TI.3:} \textit{Increasing production quality}). Furthermore, while some of their participants argued that (3) their creative process is too complex for AI to imitate, we only observed a similar pattern on the level of the entire production, e.g. highlighting shortcomings in bringing everything together (\link{TI.4.3:} \textit{To fit art into big picture}). We are surprised that their participants critique creative AI's value in communicating with clients: \enquote{Often we can do that better with a scribble than a fancy looking piece of art}. We hold that this observation is grounded in specific industry practice, in that the reverse has been explicitly stressed by our participants (\link{TI.4.2}). While about half of their respondents expressed strong excitement about the contribution of creative AI to their profession, this theme is arguably even more expressed in our data (\link{PS.1:} \textit{Excited about systems}). We can moreover confirm all of the associated sub-themes, namely that (1) AI can increase productivity (\link{TI.3:} \textit{ Increasing efficiency}), (2) offer inspiration (\link{US.5:} \textit{As a source of inspiration (a priori)}), and (3) lead to higher quality output (\link{TI.3:} \textit{Increasing production quality}).}

\section{Discussion}
\label{sec:reflection}





\rev{What are the implications of our findings? They confirm that \ac{TTIG} has firmly arrived in game industry, and will likely impact the creative process all the way from early ideation to assets used in the final product. Most importantly, our findings identify the stakeholders which shape this development, and make both opportunities and threats transparent. To only name a few, \ac{TTIG} could empower small studios, create new jobs, enable richer or entirely new types of games, and alleviate artists from repetitive creative work while supercharging their agency and ability for creative expression. However, \ac{TTIG} could also cost existing jobs, deny new and extensively trained workforce entry to the industry, yield more but not better games, cause guilt through mandated use of AI against individual ethics, and promote the loss of meaning and agency in artistic work, ultimately impeding well-being on a large scale. Arguably our most important conclusion is that everything is in flux, and we know too little to inform responsible decisions. Iterating a sentence from the introduction, we see ourselves at a crossroads. We now have a better idea where we could end up, but so far, we only have little insights to guide us towards the right direction. To serve us as a compass, we ask: what does it require for creative \ac{AI} to benefit all of us, and not only a few? We refrain from giving many recommendations for specific action, but instead promote a mindset emphasising democratic principles, empathy and collaboration.}

\rev{We want to stress that most of our interviewed game professionals already expressed this mindset. They were eager to share their thoughts, and already engage heavily in discussions and critical thinking. Game industry has a tradition of adopting new tools, and professionals are accustomed to change \cite{kultima2018game, kerr2017global}, as can be seen e.g. in our participants' pragmatic views. Moreover, they have typically strong self-awareness of their own impact, cultivated by a tradition of taking player feedback into consideration. In recent years, game industry has dedicated more attention to issues such as work-life-balance, crunch / studio culture, leadership styles, and diversity in teams, approaching these through a democratic lens. Games industry communities are no more mere vessels of technological determinism, and the level of opportunistic views are lowered (but still present, cf. our title quote). Having become more critical on the shifts that their practice goes through, we deem game professionals keen to embrace the sustainable adoption of this new technology. However, their efforts are hampered, e.g. by ethical concerns (especially copyright) and fear about the implications for art, craft, and jobs. To resolve this, we argue that they must also rely on, and demand contributions from other stakeholders, as listed below. We consequently encourage game professionals to carry their discussions even more strongly to the outside, make themselves heard, and become active participants in shaping this technology and its adoption. Next to a democratic attitude, our data also indicates much empathy for fellow artists and a strong appreciation of the human element in game production. Crucially though, it is yet to be seen to which extent e.g. hiring decisions in favour of people can withstand commercial pressures. Many of these are exercised by the players as customers, which is why we encourage companies to keep exploring the use of \ac{TTIG} for the production of games that were not possible to realise before, e.g. with increased richness and diversity of content, or entirely new game designs. Based on the opinions of our interviewed professionals, it is these games that have the most potential to sustain jobs and meaning in creative work, while leveraging this new technology to the fullest. Likewise, we encourage players to be discerning and demand games that go beyond the state-of-the-art, especially from big studios.}

\rev{As HCI researchers, we can support this development by collecting more data from professionals in a longitudinal fashion. Our study highlights the value of studying the impact of creative \ac{AI} in a more ecologically valid setting, with real users, in their natural workplace, on real-world tasks, and across the whole creative process \citep{uusitalo2022co}. The data from this and related work \citep[][cf.~Sec.~\ref{sec:related_academic}]{ko2022large,buraga2022emergence,inie2023} has been  collected within a very limited time frame, and effectively communicates immediate needs for action, e.g. on copyright legislation. However, the fast pace of technical development and change of attitudes even within the same participant also highlights its insufficiency for reliably forecasting future developments for long-term planning. In addition to collecting and analysing data, HCI researchers can use the insights to put forward design principles for AI systems to augment and complement, rather than substitute, human creativity. As an example, our data suggests that, at present, game professionals are more willing to adopt \ac{TTIG} as part of the pre-production phase (such as ideation, inspiration, communicating the ideas and, prototyping), using it in an exploratory fashion. To support this use, it seems worthwhile to promote embracing weirdness and divergence instead of mastering the visual conventions. Consequently, HCI researchers have begun asking: how can we design creative \ac{AI} systems for playfulness \citep{liapis2023designing}? Stressing democracy, empathy and collaboration, we suggest to frame these collective efforts as research on human-machine co-creativity \citep{kantosalo2016modes} with the values of participatory AI \citep{Abeba2022}, ideally moving beyond a human-centric to a more-than-human perspective on AI \citep{jaaskelainen2022exploring} to embrace sustainability beyond social factors.}

\rev{Following the same principles, we urge academic and industry AI researchers to collaborate more strongly with their HCI colleagues to responsibly balance scientific curiosity and financial incentives with the potential implications of their research on its professional users. While applied \ac{AI} research has long been focusing on automating hard, tedious, and unrewarding work, our data supports that it has now shifted into the realm of human self-realisation. Hence, new creative AI systems have the potential to alleviate opportunities for meaning-making through work, thus requiring more careful reflection. Our data can also inform AI research in that it highlights weaknesses in existing \ac{TTIG} models and professionals' requests for their future development.}

\rev{Our data highlights the urgent need for policymakers to adjust copyright law in order to fairly compensate professional creatives, relieve them from ethical concerns, and support the protection of the generated assets for use in games as commercial products. Echoing the previous paragraph, we also urge legislators to consider how such technology could not only increase productivity, but also be detrimental to the generation of meaning through jobs, with potential long-term implications for personal well-being and thus public health. Moreover, we also note that the use of \ac{TTIG} to substitute workforce or skill acquisition comes at the price of a stronger dependency on technology.}

\rev{We finally discuss implications for educators, including the authors of this paper. At present, \ac{TTIG} is adopted as an integral part of artists' larger tool sets. This in return also implies that we need to reiterate the pipeline models that are taught when educating future professionals for the creative industries, as highlighted by our participants. Moreover, we should consider how the impact of \ac{TTIG} on fulfilling artistic needs could be compensated for.}

\rev{Crucially, our data suggests that there is no one-fits-all solution, but we urge to leverage each other's excitement about this technology and listen carefully to push this technology forward.}

\section{Study Features \& Limitations}
\label{sec:limitations}

Our study has particular features and limitations that deserve attention. \rev{Extending the discussion in Sec.~\ref{sec:study_participants}, we note that} our participant sample (14) was relatively small, but robust enough for an exploratory study. To highlight the context: Finnish game industry was estimated to have employed approximately 3600 workers in 2020 \citep{hiltunen_2020} \rev{and 4100 in 2022 \citep{hiltunen_2022}}, out of which the portion of visual professionals is somewhere around 1000 (from personal correspondence). About 22\% of all employees identify as female  \citep{hiltunen_2020,hiltunen_2022}. While having a narrow diversity in representing all genders, our sample follows the regional representation. There is no detailed information of the ratio of titles and role compositions in the companies, but the median size of the companies \rev{\cite[8 in 2020, 10 in 2022][]{hiltunen_2020}} supports that Finnish game professionals \enquote{wear many hats}. This makes us confident that we also sufficiently covered different roles.

The interviews were all conducted in English, despite nine of the interviewees being native Finnish speakers. It could be argued that this impacts the depth of the views expressed due to the use of a foreign language \citep{van2010language}. While the selection of the interview language was partly dictated by practical reasons -- some of the authors not being Finnish -- we also note that most of the capital area game companies in Finland have a high ratio of non-Finnish workers (28\% nationally \cite{hiltunen_2020}). Consequently, the professional language of game industry in Finland is predominantly English. Even when practitioners discuss their work in Finnish, it is often flustered with English words, and few know the translations for the game industry lingo. The language area of Finnish is so small, that most of the games are never localised to Finnish, and the Finnish game vocabulary is mainly kept alive by the local game press. Game industry professionals working in Finland are used to expressing complicated thoughts about their work in English. No-one informed us of language struggles explicitly, and we also did not observe them implicitly in the interviews or in our data.


\rev{While our questions focused on the use of \ac{TTIG} in the workplace, we realised that especially in the adoption phase, much exploration of these systems happened outside official working hours. We however did not take into account statements about using \ac{TTIG} outside work unless the difference was clearly articulated and reflected on. This is because such use might come with different demands and constraints (e.g. time pressure, requirements for quality and consistency, etc.). At the same time though, we root our conceptualisation of professional creatives in their skill and experience, and do not want to limit it to (paid) work. Our present insights highlight opportunities in embracing a more holistic view: several participants reported the active or potential use of \ac{TTIG} in their \enquote{own projects} outside work, which can have much resemblance with their professional tasks. At the same time, others referred to such projects as an opportunity to escape dictated usage at work. Future studies should leverage more structure in interviews to capture related experience in and outside the workplace, while making the respective demands and constraints transparent.}



\rev{Our study captures the views surfacing within a particular slice of time in the adoption and development of creative AI. Any such period after the introduction of a new technology comes with its unique opportunities and challenges. Conducting this study right after \ac{TTIG} was first made available through waiting lists and invitations by peers would have allowed us to capture people's immediate, first impressions with fewer influences from a larger public debate. However, finding participants with access to, and experience in using \ac{TTIG} would have been more difficult, and introducing them to the systems ourselves might have added bias. At the time of conducting our interviews (Nov-Dec 2022), the topic of creative AI was a popular, heated, and hyped discussion piece. We hence note that our participants' views reported here might have been shaped by surrounding discussions with near and far peers, as well as more general opinions expressed in social media and the news, in addition to their own first-hand experiences and reflection.}

\rev{Due to the fast pace of \ac{TTIG}'s development and \ac{AI} more generally, our findings on the reported system strengths and weaknesses (\link{SSW}) and consequently their use (\link{US}) are highly volatile. As a specific example, one participant requested to also use images as input, complementing the text prompts (\link{FD.2}). In the meantime, this feature has not only been realised in most available \ac{TTIG} systems, but its effective use, e.g. to specify a certain desired output style, has also been investigated in academic studies \citep{qiao2022initial}. We moreover want to emphasise that people's attitudes are dynamic, shaped e.g.~through first-hand experience, industry leadership, changes in legislation and the wider public debate. This is reflected, amongst others, in how professionals perceive these systems (\link{PS}), and how they think about their impact (\link{TI}) and future development (\link{FD}). We do not consider this a shortcoming, but a feature of our study which comes to full fruition when comparing insights with further studies on these and similar phenomena in a longitudinal fashion.}

\rev{To better support comparisons with and between such future studies, we provide two recommendations informed by our shortcomings. Firstly, it would have been extremely helpful in evaluating individual statements on the capabilities (e.g.~\link{SSW}) of \ac{TTIG} systems if we had augmented the respective quotes with information on the system version used. Collecting such information is challenging, in that our participants used several different systems and versions at a time; this actually motivates the collection of such information to enable fruitful comparisons between future studies. Secondly, the quotes would afford richer interpretation given more precise information on individual expertise in using \ac{TTIG} specifically, and on \ac{AI} literacy more generally. Resting on our semi-structured interview, our present categorisation in Tbl.~\ref{tbl:participants_overview} is incomplete, and only serves as a rough approximation of expertise based on \textit{usage frequency} (\link{US.2}). We hold that future studies will benefit from collecting such data through (ideally validated) questionnaires.}

\rev{Creative \ac{AI} such as \ac{TTIG} currently impacts creative industry at large. Hence, it seems prudent to ask whether our findings, and their implications, translate beyond games. In order to not invite inappropriate and potentially harmful generalisations, we urge caution with straight-forward generalisations, and instead advocate a more nuanced view based on comparing games with other creative professions. For instance, our data has been heavily influenced by how well game professionals are accustomed to change through digital technologies, and how they communicate and collaborate toward a final product. Also, existing research shows that game industry might well be overall more techno-optimist, and has cultivated a stronger continuity in embracing tools and innovation compared to other industries \cite{kultima2018game, kultima2010hopefully, tschang2007balancing}. To pave the way towards a more general theory, further research is needed to understand how creative and collaborative processes differ between industries, paired with insight on professionals' perception, adoption and use of creative \ac{AI}. Given such additional data, we see potential in applying some of our insights to other applied art disciplines where visual prototyping and ideation plays an important role. Game industry offers insights into many types of creative practice under one roof, and could assume a leadership role in the sustainable adoption of creative \ac{AI}.}

\rev{We finally discuss how local characteristics of the Finnish game industry may impact the generalisation of our findings not only to other creative professions, but also global game industry. One important factor is that the Finnish game industry} is built on the experiences of smaller scale game development, where the role of a concept artist can be comparatively diminished. Finnish game companies are often very small (median 8, average 25; in 2020 \cite{hiltunen_2020}), favouring the mixture of roles instead of dedicated titles. While concept art and visual sketching is part of the pipeline, most of the companies do not have anyone that would be dedicated to that role only. The existence of the focused concept artists in Finland was even questioned by one of the interviewees: \enquote{(...) there's a lot of talk that, for example, concept artists will lose their jobs, but the the fact is that at the moment, I don't really know how many concept artists there are in the world. Or at least, If I look at the game industry in Finland. I don't even know any concept artists. We don't have any concept artists here (...)} (5, M, 34-44, Artist, +++). The discussion impacting Finnish game developers is extending over national borders, but should at the same time be interpreted within the Finnish ecosystem. 

\section{Conclusion \& Future Work}
In this paper, we have present the first study of professional creatives' perception, adoption and use of \acf{TTIG} systems in a specific industry, namely videogames. Based on a exploratory Template Analysis of 14 interviews by professional creatives in much varied roles, we establish what we believe is the most comprehensive and yet detailed collection and discussion of insights on this phenomenon to date. Portraying the plurality of views expressed by the professional creatives in Finnish game companies, it opens up a window to deeper issues underlying the public discussions on the wider adoption of TTIG tools. Working on the verge of a (yet another) change of roles and creative processes, our participants expressed excitement, concerns, ethical reflections, and commented on the use of the tools in their practice. Furthermore, the reflections of our interviewees show signs of inner conflict. 

\rev{We also contribute a discussion of how our findings can inform the actions of the most important stakeholders toward the sustainable adoption of \ac{TTIG}. Creative \ac{AI} more generally has been highlighted as one of the \enquote{major technological opportunities and challenges} in the Finnish game industry 2022 \cite{hiltunen_2022}, and we consider this statement to hold beyond national borders. We hope that this report will make game professionals feel heard, and inform further discussions about shaping a sustainable future of their work within and beyond the industry. Moreover, we are confident that our insights can alleviate uncertainty in the implementation of creative \ac{AI} in game development pipelines and inform AI-driven business strategies, two of the challenges raised in the latest industry report \cite{hiltunen_2022}. Our study provides richer, more reliable, and industry-specific insights for policymakers} to inform urgently needed legislation supporting the ethical use of creative AI, as also demanded by our participants. Finally, our insights can inform researchers in games, HCI and AI to advance the sustainable use of creative AI, and benefit games as cultural artefacts.

We highlight the most important avenues for future research, as motivated in the previous two sections. Our focus on industry professionals leaves out the voices of students, i.e.~those who have already received professional training, but have not joined the industry yet. This particularly vulnerable demographic, with a different skillset and experience but also subjected to different pressures, should be included in future work. Moreover, our study only marks a snapshot on professionals perception, adoption and use of \ac{TTIG} at a specific time. In order to understand the development of the addressed phenomena, and provide up-to-date, actionable data, longitudinal studies across different countries are required. As part of that, we are also interested in better understanding how dynamics within the game industry, including those between artists and their team leads, influence the views on, and adoption of, \ac{TTIG}. \rev{We have provided recommendations on how to make such future studies most effective, and support longitudinal comparisons. Given the fast pace of development, we invite our readership to comment on our findings against their subjective experience.}



\begin{acks}
We thank all participants for their time and for sharing their thoughts and attitudes with us. \rev{We also thank our reviewers for their excellent feedback and encouragement.} Moreover, we are grateful for inspiring discussions with André Holzapfel. Petra Jääskeläinen, Anna-Kaisa Kaila from the KTH project \enquote{AI and the Artistic Imaginary}. Moreover, many thanks go to Minna van Gerven for helpful discussions on Thematic Analysis. The first author of this work received financial support by the Helsinki Institute for Information Technology (HIIT, grant no. 9125064).
\end{acks}

\appendix

\section{Details on Our Template Analysis}
\label{sec:appx_analysis}
We study our data through Template Analysis \citep{brooks2015utility}, a specific form of Thematic Analysis - a collection of qualitative data analysis methods that \enquote{seek to define themes within the data and organise those themes into some type of structure to aid interpretation}. The benefits of Thematic Analysis have been well summarised by \citet{braun2006using}; we use it specifically for its potential to generate unanticipated insights, highlight similarities and differences across the data set, invite social as well as psychological interpretations, and yield a thick description of the data set that is accessibility to an educated general public, and particularly suitable for informing policy development. We chose Template Analysis \citep{brooks2015utility} over other methods for three specific features. Firstly, it emphasises the development of a coding template early on, based on a subset of the data. This template, describing a hierarchy of codes and themes, is then applied to further data, revised and refined. The use of a template allowed us to collaborate more easily on the analysis, and to handle the large amounts of data from our interviews. Secondly, the initial template may comprise a priori codes and themes. This enabled us to include the topics identified in Sec.~\ref{sec:related_work} and thus connect our findings to the existing, albeit more shallow, discussion in related work. Thirdly, Template Analysis encourages coding hierarchies with more than two levels, allowing us to capture richer and more detailed aspects of our data. 

To make our epistemological assumptions transparent, we reflect on the types of decisions that shaped our analysis as laid out by \citet{braun2006using}. 
We set out to not only capture one facet, but to devise a \emph{detailed account of multiple aspects of the data set}. This marks a compromise between exploring important themes in the data which might not relate to our overarching research questions (Sec.~\ref{sec:introduction}), while also supporting the latter. Our approach is mostly \emph{inductive}, in that the majority of the identified themes are closely linked to the data itself. At the same time, the structuring of the interviews based on our research questions, together with the a priori themes included from related work (Sec.~\ref{sec:related_work}), imply that a small part of the codebook structure has been influenced by existing -- albeit narrow and shallow -- theory, and the researchers' interests. We chose to identify themes primarily at a \emph{semantic} level, i.e.~based on the surface meaning of what was said, and only employed a \emph{latent} approach in a few instances where the intended meaning was obfuscated but obvious for all coders. 
Our guiding research epistemology is \enquote{subtle realist} \citep{brooks2015utility}, in that we believe in our ability to identify phenomena independently of our researcher identity, and yet acknowledge that the latter, including our proximity to our demographic and their work, inevitably influences our interpretation; our research questions represent a \enquote{matter of care} \citep{de2011matters}, in that we identified them from a position of personal empathy and curiosity. 

The Template Analysis was conducted by two of the authors with the Atlas.ti qualitative data analysis software, adopting the six steps defined by \citet{brooks2015utility} to enable asynchronous, collaborative coding. We first familiarised ourselves with the data by independently reading five randomly selected transcripts (step 1). We then carried out preliminary and independent coding of one randomly selected, non-overlapping transcript (step 2). Here, we already used the a priori codes from related work (suffix \enquote{(a priori)} in the codebook). We chose to code fragments ranging from partial sentences to at most one paragraph, and contributed memos \citep{crabtree1992template} to a shared document when spotting interesting relationships between codes and themes. Once we completed our own coding, we reviewed our collaborator's transcripts independently, adding codes, and noting suggestions for changes. After this first round of coding, we consolidated our ideas, and defined the initial codebook by organising all codes into a hierarchy of themes, sub-themes, etc., revisiting the coded fragments where necessary to refine our clustering and naming (steps 3 \& 4). We then repeated this process of independent coding based on the latest template, swapping, reviewing and consolidating until all transcripts were coded (step 5). Since most new codes and themes were formed early in the process with the codebook converging along the way, we successively increased the number of coded transcripts, allowing us to dedicate roughly equal amounts of time to new additions in each consolidation session. Based on an additional round of two, then moving to four and finally six transcripts between the coders, we produced codebook versions two to four. To determine our final coding template (step 6), we once more reviewed all codes and themes, added sub-themes where a theme contained too many and/or heterogeneous codes, and deleted codes / themes with too little grounding. In this process, we weighed the insights that a code contributed against their prevalence; in some instances, we kept a code even if it was only grounded in one interview, because it voiced a very different view, or closely related to a similar code in another theme. In the end, all transcripts were coded at a high density, except for very few parts unrelated to \ac{TTIG} systems. This indicates that the template development has converged \citep{brooks2015utility}. Coding one transcript took about 3 hours, reviewing it 2 hours, and each consolidation session lasted about 3 hours. All codebook versions are provided in the Supplementary Materials. 

\rev{
\section{Semi-structured Interview Outline}
\label{sec:appx_interview}
The interview was designed for approximately 60 minutes, and we divided it into five segments (Background/Warming up, Perception of the systems, Use of the systems, Future of use, Wrap-up). Each segment has been defined in terms of primary questions to ask (latin numerals) and sub-questions (roman numerals) to pose where appropriate. 

\paragraph{Background/Warming up (5-10 min)}
\begin{enumerate}
\item Current role?
    \begin{enumerate}
    \item What are you working on right now?
    \item What are the most common tasks in your day-to-day job?
    \end{enumerate}
\item Work history?
    \begin{enumerate}
        \item How long have you been working as a game designer/artist/developer? In
which roles?
    \end{enumerate}
\item What do you think of when you hear the term “AI image generators”?
\item When were you first introduced to one of the systems?
\begin{enumerate}
    \item What were your first thoughts?
\end{enumerate}
\item Your experience with using text-to-image generators
\begin{enumerate}
    \item How do you use them? In what context?
    \item How much? (never, once, several times, weekly, daily)?
    \item Which tools?
\end{enumerate}
\item Do you work with other AI in your daily practice (outside image generation)
\end{enumerate}
    
\paragraph{Perception of the systems: (10 min)}
\begin{enumerate}
\item What kind of impact do you think these systems will have on your work?
    \begin{enumerate}
    \item If not already discussed, also ask: If you think about the game industry beyond your own work, what kind of impact will these systems have?
    \end{enumerate}
\item What value do you think these systems bring?
    \begin{enumerate}
    \item Can you think of any examples?
    \end{enumerate}
\item Do you have any concerns related to these systems?
    \begin{enumerate}
    \item Can you think of any examples?
    \end{enumerate}
\end{enumerate}
    
\paragraph{Use of the systems (5-15 min depending on their level of experience)}
\begin{enumerate}
\item How would you describe the creative process when using one of these
systems?
    \begin{enumerate}
    \item How does the creative process change with these tools?
    \item What is the difference between pastime/professional use? Is there any?
    \end{enumerate}
\item What motivates you to use these systems?
    \begin{enumerate}
    \item What are the systems good for?
    \item What are your best experiences with using the systems?
    \item What do you look for in the results?
    \item Have you noticed any patterns/best practices? What works? What doesn’t?
    \end{enumerate}
\item Can you identify any challenges in using the systems?
    \begin{enumerate}
    \item What doesn’t work?
    \item What are your worst experiences/disappointments with using the systems?
    \end{enumerate}
\item Is there something that stops you from using the systems?
    \begin{enumerate}
    \item Do you think you have the skills and knowledge to use these systems? 
    \item Where do you find the knowledge/learn how to use these systems? 
    \item Do you think your background as an artist helps you in using these systems?
    \item Do you have all the necessary resources for using the systems?
    \end{enumerate}
\end{enumerate}

\paragraph{Future of use (15 min)}
\begin{enumerate}
\item Do you see yourself using these systems in the future (in your work)? Why?
    \begin{enumerate}
    \item If not already discussed, also ask: If you think about the game industry beyond your own work, do you see these systems being used in the future? Why?
    \end{enumerate}
\item Do you think it will be necessary to properly know how to use these systems in your work?
    \begin{enumerate}
    \item Will it be an integral part of your role?
    \item How do you see your role changing?
    \item How much do you think you can decide for yourself on the use of the systems?
    \end{enumerate}
\item What kind of tasks do you see yourself “outsourcing” to a text-to-image generator?
    \begin{enumerate}
    \item How do you feel about this? Positives/negatives?
    \end{enumerate}
\item Art outsourcing is common in the game industry. How do you think outsourcing image creation to AI differs from outsourcing to humans working outside your own company?
\item What do you expect from systems in the future?
    \begin{enumerate}
    \item What should the systems be better at?
    \end{enumerate}
\end{enumerate}

\paragraph{Wrap-up (5 min)}
    \begin{enumerate}
    \item If you could pick one thing people should know about these systems, what
    would it be?
    \item Is there anything else that you would still like to add?
    \end{enumerate}

}

\bibliographystyle{ACM-Reference-Format}
\bibliography{bibliography}

\end{document}